\documentclass[%
 aip,
 amsmath,amssymb,
reprint, onecolumn %%%%% Comment onecolumn to override one column format %%%%%%%
]{revtex4-2}
\usepackage{amsmath, amssymb, amsfonts, braket, commath, bm, empheq, array, lipsum}

\usepackage[]{graphicx}
\DeclareGraphicsExtensions{.eps,.jpg,.png,.pdf}
\usepackage{hyperref, url}
\hypersetup{
	colorlinks=true,
	linkcolor=red,
	filecolor=blue,
	urlcolor=magenta,
}

\usepackage{dblfloatfix}

\usepackage{xcolor}
    \definecolor{darkblue}{rgb}{0,0,.65}
    \definecolor{darkgreen}{rgb}{0.3,0.6,0.3}
    \definecolor{darkorange}{rgb}{0.85,0.65,0.3}
    \definecolor{cyan1}{rgb}{0.0, 0.6, 0.6}
\usepackage{soul}
\usepackage[normalem]{ulem} % needed for \sout-command

\newcommand{\bte}{$\beta$~ensemble~}

\newcommand{\Capheat}{\ensuremath{\mathcal{C}}}

\newcommand\EM{\ensuremath{E_\mathrm{ME}}}

\newcommand\FC{ \ensuremath{F} }

\newcommand\iiserk{Department of Physical Sciences, Indian Institute of Science Education and Research Kolkata, Mohanpur 741246, India}

\newcommand\lboro{School of Science, Loughborough University, Loughborough, Leicestershire LE11 3TU, United Kingdom}

\newcommand\meanT[1]{\ensuremath{\overline{#1}}}
\newcommand\mean[1]{\ensuremath{\left\langle #1 \right\rangle}}

\newcommand{\nrm}{\ensuremath{Z}}

\newcommand{\prob}[1]{\ensuremath{\textrm{P}\del{#1}}}
\newcommand\phaseA{annealed}
\newcommand\phaseQ{quenched}
\newcommand\phaseF{frozen}

\newcommand\SMC{ \ensuremath{\mathcal{S}} }
\newcommand{\sff}[1]{\ensuremath{\mathcal{K}\del{#1}}}
\newcommand{\sffdc}[1]{\ensuremath{\mathcal{K}_\mathrm{dc}\del{#1}}}
\newcommand{\msffdc}{\ensuremath{\mean{\mathcal{K}_\mathrm{dc}}}}
\newcommand{\sffc}[1]{\ensuremath{\mathcal{K}_\mathrm{c}\del{#1}}}
\newcommand{\msffc}{\ensuremath{\mean{\mathcal{K}_\mathrm{c}}}}
\newcommand\sffbar{ \ensuremath{ \overline{\mathcal{K}} } }
\newcommand{\sgE}{\ensuremath{\sigma_{E}}}
\newcommand\tdip{\ensuremath{t_\mathrm{dip}}}
\newcommand\tH{\ensuremath{t_\mathrm{H}}}
\newcommand\tR{\ensuremath{t_\mathrm{R}}}
\newcommand\tZ{\ensuremath{t_\mathrm{Zeno}}}

\newcommand\TM{\ensuremath{T_\mathrm{ME}}}

\newcommand\Tc{\ensuremath{T_c}}
\newcommand\Tq{\ensuremath{T_q}}

\newcommand\tauR{\ensuremath{\tau_\mathrm{R}}}

\begin{document}
%============================================================
\title{Thermodynamics and finite temperature spectral form factor of the $\beta$ ensemble}

 \author{Basudha Roy}\email{br23rs008@iiserkol.ac.in}
 \affiliation{\iiserk}
 \author{Adway Kumar Das}\email{A.K.Das@lboro.ac.uk}
 \affiliation{\lboro}
\author{Ivan M.\ Khaymovich}\email{ivan.khaymovich@gmail.com}
\affiliation{Univ. Toulouse, CNRS, Laboratoire de Physique Th\'eorique, Toulouse, France}
 \author{Anandamohan Ghosh}\email{anandamohan@iiserkol.ac.in}
 \affiliation{\iiserk}
%\author{authors}
%\date{}
%============================================================
%	ABSTRACT
\begin{abstract}
In this work, we study the thermodynamics and the finite-temperature spectral form factor (SFF) of the Gaussian \(\beta\) ensemble, where the level-repulsion parameter scales with system size as \(\beta=N^{-\gamma}\). By studying the system-size scaling of the ground-state Gibbs weight, we identify three thermal regimes: a frozen phase, a quenched phase, and an annealed phase. In the frozen phase, the Gibbs weight is concentrated on the ground state, and the quenched and annealed averages of the thermodynamic quantities become inequivalent. In the quenched phase, the Gibbs distribution spreads over an extensive but vanishing fraction of the spectrum, and ensemble equivalence is restored, although equipartition is still absent. In the annealed phase, the Gibbs weights become approximately uniform for a finite fraction of the total number of energy states. We show that the temperature scale separating the frozen and quenched phases depends on the system parameter \(\gamma\), whereas the scale separating the quenched and annealed phases is independent of \(\gamma\) for a fixed spectral bandwidth. We further study the finite-temperature SFF and analyze the system-size scaling of its characteristic timescales and the correlation-hole depth in the nonergodic extended (NEE) regime, and verify the absence of mobility edge (ME) within this regime. Finally, we use the finite-temperature Lanczos method (FTLM) to calculate the SFF without requiring the full spectrum specifically at low temperatures and show that its temperature dependence can be used to estimate the infinite-temperature relaxation time.
\end{abstract}
%===========================================================
\maketitle

%===========================================================
\section{Introduction}

In quantum systems, spectral correlations play a crucial role in identifying the chaotic nature of the system. Particularly, in completely chaotic/ergodic quantum systems~\cite{Bohigas1984}, the energy levels exhibit level repulsion, such that they avoid being too close to each other, and their statistics are well described by the random matrix theory (RMT)~\cite{Dyson1962,Mehta2004book,Forrester2010,Guhr1998,Cotler2017a}. In contrast, the energy levels in integrable/localized quantum systems follow Poisson statistics~\cite{Berry1977, Anderson1958}, and remain uncorrelated at all length scales. Between these two limits lies another class of systems that has attracted considerable attention because of the nonergodic extended (NEE) nature of their eigenstates, i.e., the eigenstates are spatially extended but occupy only a vanishing fraction of the available Hilbert space. Such states have been studied in the Rosenzweig--Porter model~\cite{Kravtsov2015}, banded random matrices~\cite{Das2025}, the $\beta$ ensemble~\cite{Das2022, Das2023}, and other models such as the quantum random energy model~\cite{Faoro2019} and the Sachdev-Ye-Kitaev model~\cite{Micklitz2019}. The specific model with a NEE regime we are interested in here is the Gaussian $\beta$ ensemble~\cite{Dumitriu2002}, a random matrix model with explicit control over energy correlations. The tridiagonal matrix representation of this ensemble~\cite{Dumitriu2002} reveals a connection with random discrete Schrödinger operators~\cite{Breuer2007}, whose spectral properties exhibit different regimes depending on the Dyson index $\beta$. This analysis was further extended in~\cite{Breuer2010}, where random Jacobi matrices with power-law growing off-diagonal elements were studied with applications to the $\beta$ ensemble.

The correlations at different energy scales are reflected in the time evolution of the spectral form factor, which is related to the Fourier transform of the two-level correlation function~\cite{Mehta2004book,Brezin1997,Guhr1998, Cotler2017}. For chaotic systems, it exhibits the familiar dip--ramp--plateau structure~\cite{Cipolloni2023,Winer2022}. The ramp and the correlation hole, defined as the dip in the time evolution of an observable below its saturation value, are consequences of spectral rigidity and level repulsion~\cite{TorresHerrera2017}.

At infinite temperature, every energy level contributes equally to the SFF. However, at finite temperatures, the Gibbs distribution assigns unequal weights to different parts of the spectrum. Thus, the finite-temperature SFF which equals the survival probability of the coherent Gibbs state~\cite{DelCampo2017,DelCampo2018,Xu2021}, connects dynamical, spectral, and thermodynamic information within a single quantity. This connection can be exploited to identify a mobility edge from the finite-temperature SFF~\cite{Roy2026}. As the temperature is lowered, the Gibbs distribution progressively selects the lower part of the spectrum. It is therefore natural to ask how the correlation hole and the characteristic timescales of the SFF change with temperature, and whether their temperature dependence is consistent with their infinite-temperature limits~\cite{Roy2025}.

Introducing temperature into the model makes it particularly interesting to study its thermodynamics and understand the corresponding changes in its dynamical features. In studying the thermodynamics, we show that the $\beta$ ensemble has three thermal regimes: a frozen phase, in which the Gibbs weight is concentrated on the ground state and the distinction between quenched and annealed averages becomes important~\cite{Baldwin2020}; a quenched phase, in which the coherent Gibbs state spreads over an extensive but vanishing fraction of the eigenstates and ensemble equivalence is restored; and an annealed phase with approximately uniform Gibbs weights. In particular, the boundary between the frozen and quenched phases is nontrivial because it is determined by the fluctuations of the ground-state energy, which depend on the system parameters. In contrast, the transition from the quenched phase to the annealed phase essentially depends on the spectral bandwidth. We further compare the temperature dependence of the correlation hole with the energy-resolved inverse participation ratio~\cite{Das2023}. This comparison allows us to examine whether the finite-temperature SFF can distinguish different spectral regimes and, in particular, to investigate the absence of a ME in the NEE phase.

An important practical limitation in studying the SFF is that its direct evaluation requires the entire energy spectrum. This makes exact diagonalization computationally expensive as the system size grows, particularly when one needs to extract the scaling of the characteristic timescales. To overcome this limitation, we use the fact that the finite-temperature SFF can be expressed in terms of the analytic continuation of the partition function to complex temperature~\cite{Cotler2017a,Xu2019,DelCampo2020}. This partition function can be estimated without explicitly calculating the full spectrum, using Krylov-space methods. In particular, the finite-temperature Lanczos method~\cite{Jaklic1994, Prelovvsek2013}, allows us to approximate the SFF using only a small number $m\ll N$ of Lanczos steps at low temperatures~\footnote{Similarly, to the calculation of the return probability in the fractal phase or in the critical points, see, e.g.,\cite{Jiricek2026}}. Related approaches include the kernel polynomial method~\cite{Weisse2006,Sierant2020} and stochastic Lanczos quadrature~\cite{Ubaru2017, Li2022}. At low temperature, the suppression of high-energy states makes the calculation particularly efficient, since relatively few Lanczos steps can provide an accurate estimate of the SFF. Furthermore, the temperature dependence of the scaling exponents derived in this work provides a way to extrapolate the finite-temperature results to the infinite-temperature limit. This approach allows us to determine the infinite-temperature timescales without calculating the entire spectrum.

The organization of the paper is as follows. Sec.~\ref{sec:model} introduces the $\beta$ ensemble and briefly reviews its spectral and eigenstate properties. In Sec.~\ref{sec:tdyn}, we study the ground-state Gibbs weight and use its scaling with system size to characterize the thermal regimes. Section~\ref{sec: sff_finite_T} focuses on the finite-temperature SFF, where we discuss its characteristic timescales, the correlation hole, and their relation to the energy-resolved IPR. In Sec.~\ref{sec:lanczos}, we describe the finite-temperature Lanczos approach and its use in extrapolating the results to infinite temperature. Our concluding remarks are given in Sec.~\ref{sec:conclusion}. Further thermodynamic results and technical details are provided in the Appendices.

%===========================================================
\section{Model: $\beta$ ensemble}
\label{sec:model}
The Gaussian $\beta$ ensemble is a one-parameter generalization of the Wigner-Dyson ensembles, characterized by the joint probability distribution of eigenvalues
\begin{equation}
    P_\beta(E_1,\ldots,E_N)
    =\frac{1}{\mathcal{N}_\beta}
    \exp\left(-\frac{1}{2}\sum_{i=1}^N E_i^2\right)
    \prod_{i<j}|E_i-E_j|^\beta ,
    \label{eq:beta_jpdf}
\end{equation}
where the Dyson index $\beta>0$ controls the strength of the eigenvalue repulsion~\cite{Dyson1962,Mehta2004book,Forrester2010} and $\mathcal{N}_\beta$ is the normalization constant.
The values $\beta=1,2,4$ correspond to the Gaussian orthogonal, unitary and symplectic ensembles, respectively.
For arbitrary $\beta>0$, Eq.~\eqref{eq:beta_jpdf} is realized by random symmetric tridiagonal (Jacobi) matrices~\cite{Dumitriu2002},
\begin{equation}
H_\beta=
\begin{pmatrix}
a_1 & b_{N-1} & & & \\
b_{N-1} & a_2 & b_{N-2} & & \\
 & \ddots & \ddots & \ddots & \\
 & & b_{2} & a_{N-1} & b_{1}\\
 & & & b_{1} & a_N
\end{pmatrix},
\qquad a_i\sim\mathcal{N}(0,1),\quad b_k\sim\chi_{\beta k} / \sqrt{2},
\label{eq:tridiag}
\end{equation}
with all entries being independent and $\chi_{\beta k}$ is the chi distribution with $\beta k$ degrees of freedom.
The tridiagonal form makes $\beta$ a continuous and tunable parameter. The above tridiagonal representation provides the route to rigorous edge asymptotics of the $\beta$ ensemble, namely the stochastic Airy operator and Tracy--Widom-type edge statistics for general values of $\beta$~\cite{Edelman2007,Ramirez2011,Tracy1994}.

Following Ref.~\onlinecite{Das2022}, we parameterize the Dyson index as $\beta=N^{-\gamma}$, so that $\gamma$ interpolates between strongly correlated and effectively uncorrelated spectra.
For $\gamma=0$ the model reduces to the Gaussian orthogonal ensemble.
For $\gamma<0$ the level repulsion grows with $N$, and the density of states (DOS) approaches the Wigner semicircle in the thermodynamic limit, with Tracy--Widom fluctuations at the spectral edge.
For $\gamma>0$ the off-diagonal elements are progressively suppressed relative to the diagonal ones. The bulk eigenvalues of the \bte roughly scale with system size as $\epsilon_\gamma=\sqrt{4+2N^{1-\gamma}}$~\cite{Das2022}.
In the subsequent numerical analysis we rescale the spectrum and all the energy variables by $\epsilon_\gamma$, such that $E'=E/\epsilon_\gamma$, thus, the spectral energy standard deviation is fixed at $1/2$.

The eigenvector and spectral properties of this model have been characterized in detail~\cite{Das2022, duy2016, Roy2025, Das2023}.
For $0<\gamma<1$ the eigenstates are nonergodic extended (NEE) with a typical fractal exponent, $D_2 = 1-\gamma$, that varies continuously with $\gamma$ and lies strictly between the localized ($D_2 = 0$) and ergodic ($D_2 = 1$) limits~\cite{Das2022, Das2023}. Here $D_2$ is extracted from the spectral average $\mean{\log I_j} \simeq - D_2 \log N$, $I_j = \sum_{n = 1}^N |\Psi_j(n)|^4$ being the inverse participation ratio (IPR) of the eigenstate at the energy $E_j$ having $n$th component $\Psi_j(n)$.
For $\gamma>1$ the eigenstates are localized~\cite{duy2016}. In the NEE phase, $0<\gamma<1$, localized states coexist with the extended states within an energy window $(-E_G', E_G')$ around the zero energy where $E_G'=E_G/\epsilon_\gamma \sim N^{-\frac{1-\gamma}{2}}$.

As shown in Refs.~\onlinecite{Roy2025, Das2023}, in the NEE phase, the spatially local mapping of the \bte to the set of the 1D Anderson models, with the model-dependent sizes $N_l$ and hopping amplitudes, leads to nearly independent spatial blocks $\Delta_l$ (which further split into sub-blocks with correlated
energies), with local mean level spacings $\delta_l$. This overlap of the spectra from different (nearly independent) blocks makes the global level spacing parametrically smaller $\delta\ll \delta_l$. Thus, in terms of level correlations, only at the energy scales $\Delta E\gtrsim \min_l \delta_l \equiv E_c$ two energy levels, separated by $\Delta E$, may come from the same spatial sub-block and become correlated and below this energy scale, $\Delta E \lesssim E_c$, the energy levels remain uncorrelated. Hence, inhomogeneous hopping terms give rise to a critical energy scale $E_c$ in case of the $\beta$ ensemble.

In the \bte the level statistics deviate from the Wigner--Dyson form once $\gamma>0$~\cite{Das2022}. Despite this nonergodicity, the spectral correlations remain strong, in the sense that the model shows partial long-range energy correlations and a correlation hole in the infinite temperature spectral form factor (SFF)~\cite{Roy2025}, which is the Fourier dual picture of the energy correlation structure. The SFF was introduced in the random-matrix context in Ref.~\onlinecite{Brezin1997} and is now a standard probe of spectral correlations in chaotic systems~\cite{Guhr1998,Cotler2017}. In the NEE regime of $\beta$ ensemble, energy correlations are anomalous in contrast to typical many-body systems, where short-range energy correlations are present. In particular, the absence of energy correlations below the critical energy scale $E_c$ is reflected in the equilibration of the SFF well before the Heisenberg time $\tH$, defined as the inverse of the mean level spacing at infinite temperature, giving rise to a temporal window termed the correlation void. The critical energy scale $E_c$ scales with the system parameter as $E_c \sim N^{\gamma-1}$~\cite{Note}. The characteristic SFF time scales and the depth of the correlation hole follow system-size scaling laws with $\gamma$-dependent exponents rather than the universal random-matrix ones.

The thermodynamic properties of the model, including the behavior of the Gibbs weight of the ground state, freezing transition, equipartition of energy eigenstates and the finite-temperature spectral form factor (SFF), are the subject of the present work together with an indication of the nonanalyticity of the free energy and ensemble equivalence. By analyzing these quantities, we identify three distinct thermal regions. In the \(\phaseF\) phase, the Gibbs weight is concentrated in the ground state, leading to a time-independent SFF and the breakdown of ensemble equivalence. In the intermediate \(\phaseQ\) phase, the coherent Gibbs state spreads over an extensive number of energy eigenstates, while the Gibbs weights remain non-uniform. In this regime, ensemble equivalence is restored, although equipartition is still absent. Finally, in the \(\phaseA\) phase, the Gibbs weights become approximately uniform across the spectrum, corresponding to equipartition and a homogeneous spreading of the coherent Gibbs state over the energy eigenbasis. Our results establish a consistent connection between the thermodynamic properties and the equilibrium spreading of the coherent Gibbs state.

To identify the characteristic temperature scales associated with the transition between the thermodynamic phases in the $\beta$ ensemble, we work within the canonical ensemble and examine suitable thermodynamic quantities.

%===========================================================
\section{Gibbs weight of the ground-state energy}\label{sec:tdyn}
%===========================================================
In a system with $N$ discrete energy levels $\{E_n\}$, the equilibrium probability of finding the system in the $n$th energy eigenstate at temperature $T$ is given by the Gibbs weight,
\begin{align}
    p_n(T) \equiv \frac{e^{-\frac{E_n}{T}}}{Z(T)},
    \qquad \sum_n p_n(T)=1,
    \label{eq:pn_def}
\end{align}
where $Z(T) \equiv \sum_{n=1}^{N} e^{-\frac{E_n}{T}}$ is the partition function. We sort the eigenvalues in ascending order $E_1\leq E_2\leq\cdots\leq E_N$, such that $p_1\geq p_2\geq\cdots\geq p_N$ and \(p_1\) is the largest Gibbs weight. The freezing transition is characterized by the dominance of the ground state in the Gibbs distribution. It is therefore natural to examine the ground-state Gibbs weight, \(p_1\), to determine the temperature separating the frozen and the quenched phase. Thus its scaling also provides a useful diagnostic of equipartition. In particular, if \(p_1\sim \mathcal{O}(N^{-1})\), all Gibbs weights are bounded from above by a quantity of order \(N^{-1}\). Together with the normalization \(\sum_{n=1}^{N}p_n=1\), this implies that a finite fraction of the total number of states must carry Gibbs weights of order \(N^{-1}\) [see Appendix~\ref{sec_appnd_p1_1/N}], which we take as the defining property of equipartition. 
%\AKD{At sufficiently low temperature, the ground state dominates the Gibbs distribution and the equipartition breaks down. We denote the region of temperature where $p_1 \sim \mathcal{O}(1)$ as the frozen phase, observed below a critical temperature $T_c$.}
%In particular, if \(p_1\sim \mathcal{O}(N^{-1})\), all the Gibbs weights are bounded from above by \(N^{-1}\). \AKD{We define equipartition of the energy spectrum as the condition where a} finite fraction of all the eigenstates carry Gibbs weights of order \(N^{-1}\).

In order to characterize the behavior of $p_1$ across different temperature, we look at its disorder average
\begin{align}
    \mean{p_1} \equiv \mean{\frac{1}{Z(T)}e ^{-\frac{E_1}{T}}}.
\end{align}
At sufficiently low temperature, the partition function may develop a broad distribution~\cite{Gardner1989} instead of concentrating sharply around its typical value. In this case, the self-averaging property is lost, i.e.~the disorder average of an observable do not capture its typical behavior and the ensemble equivalence is also absent, i.e.~the expectation value of an observable with respect to the microcanonical and the canonical ensembles do not agree~\cite{Binder1986, Mezard1986, Baldwin2020}. As a result, the disorder average, $\mean{p_1}$, is significantly affected by the rare realizations in the frozen phase. We therefore consider the typical value, $p_1^\mathrm{typ}=\exp(\mean{\ln p_1})$, %thermodynamically, 
which provides a measure of the ground-state condensation of a typical sample that is insensitive to rare samples.

Note that the disorder averaged DOS
\begin{align}
    \mean{\rho(E)}=\mean{\frac{1}{N}\sum_{n=1}^{N}\delta(E-E_n)},
\end{align}
in case of the $\beta$ ensemble for $\gamma < 1$ follows the Wigner's semicircle law in the thermodynamic limit ($N\to \infty$)
\begin{align}
    \mean{\rho(E)}=\frac{\sqrt{4\sigma_E^2-E^2}}
    {2\pi\sigma_E^2}
    \Theta(2\sigma_E-|E|),
    \label{eq:rho_E}
\end{align}
where $\Theta(x)$ denotes the Heaviside step function such that all the energy levels are within $[-2\sgE, 2\sgE]$ and $\sgE^2 \equiv \mean{E^2} - \mean{E}^2$ is the energy variance. However, for finite system sizes, the ground state energy (extreme value of a sequence of correlated random numbers) fluctuates and for $\gamma\leq 0$ follows the Tracy-Widom distribution and has $\mathcal{O}(2\sgE N^{-\frac{2}{3}})$ fluctuations around the spectral edge at $E = -2\sgE$~\cite{Dean2006, Tracy1994, Forrester2010, Mehta2004book, Shukla2022}. Then, the Gibbs weight of the ground state at temperature $T$ for $\gamma \leq 0$ is dominated by the spectral gap $E_2 - E_1$ which fluctuates in the same way as $\abs{E_1 + 2\sgE}$
\begin{align}
	\begin{split}
		p_1 &= \frac{e^{-\frac{E_1}{T}}}{\nrm(T)} \approx \del{ 1 + e^{-\frac{2\sigma_E N^{-\frac{2}{3}}}{T}}\sum\limits_{n>1} e^{-\frac{2\sigma_E+E_n}{T}} }^{-1}\\
		\Rightarrow p_1 &\xrightarrow{T < 2\sigma_E N^{-\frac{2}{3}}} 1.
	\end{split}
\end{align}
In the last line we use the fact that the above sum is at most finite $\mathcal{O}(1)$. Hence, the $\beta$ ensemble undergoes a freezing transition at 
\begin{align}
    T_c \simeq E_2 - E_1\sim 2\sgE N^{-2/3},\quad \gamma \leq 0
\end{align}
such that the system remains effectively confined to the ground state in the frozen phase, $T < T_c$. Therefore, the freezing transition is governed by the Tracy--Widom distribution for $\gamma \leq 0$. The power-law dependence of the critical temperature $T_c$ on the system size $N$ for $\gamma \leq 0$ together with the mathematically rigorous results for the energy gap $(E_2 - E_1)$, fluctuations of the ground state energy for $0<\gamma<1$~\cite{Allez2014}, and $\gamma>1$~\cite{Pakzad2019} further prompts us to look at the thermodynamics in the $\beta$ ensemble while scaling the above critical temperature as
\begin{align}
    T_c=2c_c(N,\gamma)\sgE N^{\alpha_T^c},
    \qquad
    c_c(N,\gamma)=
    \begin{cases}
    \mathcal{O}(1), & \gamma\leq 0,\\
    \mathcal{O}\big((\ln N)^{-\gamma}\big), & 0<\gamma<1,\\
    \mathcal{O}\big((\ln N)^{-1}\big), & \gamma\geq 1.
    \end{cases}
    \label{eq:alphaT_def}
\end{align}
In our numerical calculations, we scale the temperature similarly, $T=2c(N,\gamma,\alpha_T)\sgE N^{\alpha_T}$, with weak logarithmic corrections $c(N,\gamma,\alpha_T)$ which may also depend on $\alpha_T$~\footnote{Note that $c(N,\gamma,\alpha_T^c) = c_c(N,\gamma)$ while at $\alpha_T=\alpha_T^q$, defined below, $c(N,\gamma,\alpha_T^q)=O(1)$.}.
%---------------------------
%   Gibbs weight of ground state
\begin{figure}[t]
\centering
\includegraphics[width=\linewidth]{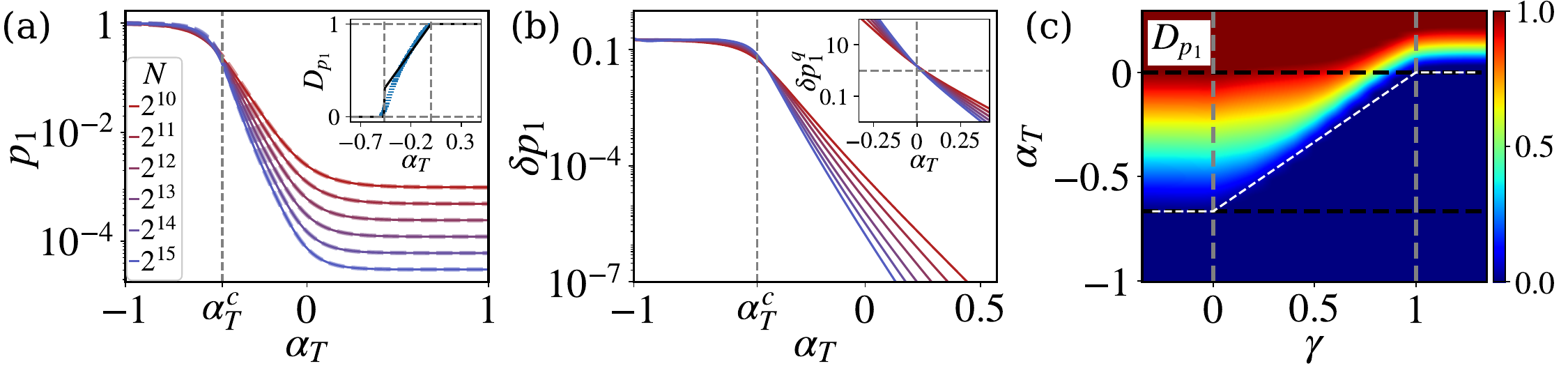}
\caption{Gibbs weight of the ground state energy of the \bte for $\sgE = 1/2$.
(a)~Average and typical values of $p_1$ shown via solid and dashed lines respectively for different system sizes and $\gamma = 0.4$. Inset shows the scaling exponent $D_{p_1} $ for the same value of $\gamma$, defined in Eq.~\eqref{eq:D_p1_def}. 
The black line shows the analytical approximation of Eq.~\eqref{eq:Dp1_Dkbar_ana}.
(b)~Relative difference $\delta p_1 = \left|1-\mean{\ln p_1}/\ln \mean{p_1}\right|$ for different system sizes and $\gamma = 0.4$. Inset shows the difference $\delta p_1^q = \left|N\mean{p_1}-1\right|$ for different system sizes and $\gamma = 0.4$ where the vertical line marks $\alpha_T^q = 0$.
%===============
(c)~Color-map of $D_{p_1} $ in the $\alpha_T-\gamma$ plane. Dashed white line marks $\alpha_T^c$ from Eq.~\eqref{eq:alpha_T_c}\, vertical grey lines indicate the range of the NEE phase.
%===============
%\BRr{(d), (e) and (f) show the difference $\delta p_1^q = \left|N\mean{p_1}-1\right|$ for different system sizes and $\gamma = -0.2$, $\gamma = 0.8$ and $\gamma = 1.2$ respectively where the vertical line marks Eq.~\eqref{eq:alpha_T^q}.}
}
\label{fig:p1}
\end{figure}
%---------------------------

In Fig.~\ref{fig:p1}(a), we show the average and typical behavior of $p_1$ as a function of $\alpha_T$ for different system sizes and a fixed value of $\gamma$. The near-common crossing of the finite-$N$ curves indicates a scale-invariant point at a critical $\alpha_T=\alpha_T^c$, where $p_1$ becomes approximately independent of $N$.
In order to compare the average and typical behavior of $p_1$, in Fig.~\ref{fig:p1}(b), we show the relative difference $\delta p_1 = \left|1-\mean{\ln p_1}/\ln \mean{p_1}\right|$ as a function of $\alpha_T$ for different system sizes. This shows that at the same value, $\alpha_T=\alpha_T^c$, the relative difference also becomes approximately independent of $N$, and the difference increases (decreases) with system size below (above) $\alpha_T^c$. This conclusively shows that the self-averaging property is absent in the frozen phase.

To identify the temperature region where equipartition of the energy spectrum is observed, we look at the following system size scaling exponent
\begin{align}
    D_{p_1}^{(N)} \equiv -\frac{\ln \mean{p_1^{(N)}}}{\ln N}.
    \label{eq:D_p1_def}
\end{align}
where $p_1^{(N)}$ is the Gibbs weight of the ground state for system size $N$ such that $\lim\limits_{N\to\infty} D_{p_1}^{(N)} \equiv D_{p_1}$.~\footnote{Numerically, we obtain $D_{p_1}^{(N)}$ for different system sizes and extrapolate with respect to~$1/\ln N$ to get $D_{p_1}$.} In the inset of Fig.~\ref{fig:p1}(a), we show the extrapolated exponent $D_{p_1} $ for the same parameter as in the main panel which confirms that $p_1$ remains finite for $\alpha_T<\alpha_T^c$ and decays towards zero with system size otherwise.

Figure~\ref{fig:p1}(c) shows the extrapolated exponent $D_{p_1}$ in the thermodynamic limit as a function of $\gamma$ and $\alpha_T$. From this quantity, we identify a critical value $\alpha_T^c$ for each $\gamma$, which is numerically found to follow the analytical predictions~\cite{Allez2014,Pakzad2019}:
\begin{align}
	\alpha_T^c \approx \begin{cases}
		-\frac{2}{3}, & \gamma \leq 0\\
		\frac{2\gamma}{3} - \frac{2}{3}, & 0 < \gamma < 1\\
		0, & \gamma \geq 1
	\end{cases}
	\label{eq:alpha_T_c}
\end{align}
below which the $D_{p_1}$ becomes zero i.e., $p_1 \sim 1$ and only the ground state survives.

At sufficiently high temperature, a finite fraction of the total number of energy states are equally probable irrespective of the nature of correlation among the energy levels, such that the average partition function becomes, $\mean{\nrm(T)} = N$. The temperature scale above which the above equipartition holds is denoted by $\Tq$ such that
\begin{align}
    T > \Tq \equiv 2\sgE N^{\alpha_T^q} \iff \mean{\nrm(T)} = N,\quad p_n \sim \mathcal{O}(N^{-1}),\quad \text{for} \quad n=1,\ldots,m -1,
\end{align}
where $\frac{m-1}{N}$ is finite. 
To find out the temperature exponent corresponding to the equiparition we consider the quantity $\delta p_1^q = \left|N\mean{p_1}-1\right|$ and show this for a fixed value of $\gamma$ in the inset of Fig.~\ref{fig:p1}(b). Numerically we find out that the curves exhibit system-size-independent crossings at $\alpha_T^q = 0$, indicating the onset of equipartition.
Note that $D_{p_1}$ also allows us to identify the temperature exponent associated with the onset of equipartition, which is $\alpha_T^q=0$ for all $\gamma$ as shown in Fig.~\ref{fig:p1}(c).
This is natural to expect that $\Tq \sim 2\sgE$ is of the order of the bandwidth, thus, for $\alpha_T\geq\alpha_T^q$ [i.e.~$T \gg T_q \sim 2\sgE$], we have $p_n\sim \mathcal{O}(N^{-1})$, indicating equipartition among the energy levels.

The freezing and the equipartition transitions are governed by different energy scales. The freezing transition is controlled by the competition between temperature and the low-energy structure of the spectrum, including ground-state fluctuations and the spectral gap. Since the low-energy structure changes nontrivially with \(\gamma\), the corresponding critical temperature acquires a \(\gamma\)-dependent scaling [Eq.~\eqref{eq:alpha_T_c}]. In contrast, equipartition requires the temperature to be large compared to the spectral bandwidth, $T>\Tq \simeq 2\sgE$, that the energy-dependent Boltzmann factors no longer produce a variation in the Gibbs weights. With the energy spectrum normalized to a fixed bandwidth, this occurs at the $\gamma$-independent temperatures.

Apart from the Gibbs weight of the ground state energy, some other thermodynamic quantities are also analyzed in Appendix.~\ref{sec_appnd_tydn}.
Combining all of the thermodynamic quantities, the \bte\ exhibits three distinct phases based on the thermodynamic quantities: frozen ($\alpha_T \leq \alpha_T^c$), quenched ($\alpha_T^c < \alpha_T \leq \alpha_T^q$) and annealed ($\alpha_T > \alpha_T^q$). The critical line between the frozen and quenched phases [Eq.~\eqref{eq:alpha_T_c}] corresponds to the freezing transition, characterized by a non-analyticity in the free energy (for $\gamma >0$) and a loss of self-averaging upon entering the frozen phase.

%===========================================================
\section{Finite temperature spectral form factor}\label{sec: sff_finite_T}
We further examine the dynamical manifestations of the thermodynamic behavior in different temperature regimes through the finite-temperature spectral form factor (SFF). The finite temperature SFF involves the analytical continuation of the partition function to the complex plane~\cite{Cotler2017a, Xu2019, DelCampo2020}
\begin{align}
	\nrm(T, it) = \sum_{i = 1}^{N} e^{-\del{\frac{1}{T} + it} E_n}.
	\label{eq:Z_cmplx_def}
\end{align}
As an entire function of the complex temperature variables ($t$ and $1/T$), $\nrm(T,it)$ is analytic throughout the complex plane of $t$ and $1/T$ and can be represented by a convergent Taylor expansion.
Its zeros, which can be accessed experimentally~\cite{Wei2012, Peng2023, Binek1998, Binek2001} are known as Fisher zeros~\cite{Fisher1965} and they encode the singular behavior associated with thermal phase transitions~\cite{Fisher1978, Kortman1971, Katsura1970, Brascamp1974, Derrida1983}.
The finite temperature SFF is given in terms of the squared modulus of the partition function
\begin{align}
	\begin{split}
		\sff{t; T} &\equiv \frac{\abs{\nrm(T,it)}^2}{\nrm(T)^2} = \frac{ \sum_{m, n}^{N} e^{i(E_m-E_n)t - \frac{E_m+E_n}{T}} }{\nrm(T)^2}
	\end{split}
	\label{eq:SFF_T_def}
\end{align}
where $t$ is the time. The finite temperature SFF can also be interpreted as the survival probability of a specific initial state called the coherent Gibbs state (CGS)~\cite{DelCampo2017, DelCampo2018, Xu2021}, $\ket{\Psi_T}$
\begin{align}
	\sff{t; T} = \abs{\langle\Psi_T\ket{\Psi_T(t)}}^2,\quad \ket{\Psi_T} \equiv \sum_{n = 1}^{N} \frac{e^{-\frac{E_n}{2T}}}{\sqrt{\nrm(T)}} \ket{\Phi_n}
	\label{eq:CGS_def}
\end{align}
where $\ket{\Phi_n}$ is the eigenstate at energy $E_n$ and $\ket{\Psi_T(t)}$ is the time evolved state. In the \phaseA~phase ($T > \Tq$), the Gibbs probability amplitude $e^{-\frac{E_n}{2T}}/\sqrt{\nrm(T)} = \sqrt{p_n(T)}$ converges to $1/\sqrt{N}$ and the $\sff{t; T}$ reduces to the infinite temperature SFF.
The initial decay of the finite temperature SFF is governed by the thermal weighted DOS
\begin{align}
	\rho_T(E) \equiv \sum_{n = 1}^{N} p_n \delta\del{E - E_n} = N \rho(E)\frac{e^{-\frac{E}{T}}}{\nrm(T)}.
	\label{eq:rho_T_CGS_def}
\end{align}
via its Fourier transform.
On the other hand, the saturation value of the SFF can be expressed in terms of the canonical partition function. Then, at sufficiently high temperature, i.e., for $\alpha_T\geq \alpha_T^q$, using the annealed approximation the disorder averaged SFF can be expressed as [see Appendix~\ref{sec_appnd_SFF_ana}]
\begin{align}
	\begin{split}
		\mean{\sff{t; T}} &\approx \abs{\int dE\; e^{-iEt}\mean{\rho_T(E)}}^2 + \del{1 - b_2\del{\frac{t}{\tH}}}\sffbar.
	\end{split}
	\label{eq:SFF_T_parts}
\end{align}
where $b_2(\tau)$ is the two-level form factor~\cite{Mehta2004book, Guhr1998}, with $\tau \equiv t / \tH$ as the dimensionless time, $\tH \sim \frac{N}{\sgE}$ as the Heisenberg time~\cite{Das2025} $\sffbar$ is the saturation value of the SFF. The first term in Eq.~\eqref{eq:SFF_T_parts} is known as the disconnected part, $\msffdc$, of the SFF whereas the second term is the connected part, $\msffc$. We now individually discuss the different parts of the finite temperature SFF.
%---------------------------
% saturation value of the SFF
\subsection{Saturation value of the SFF}
The saturation value of the SFF at finite temperature is given by
\begin{align}
	\sffbar = \mean{\frac{\nrm(T/2)}{\nrm(T)^2}}.
	\label{eq:SFF_T_bar_1}
\end{align}
%---------------------------
%   saturation value of the SFF of beta ensemble
\begin{figure}[t]
\centering
\includegraphics[width=\linewidth]{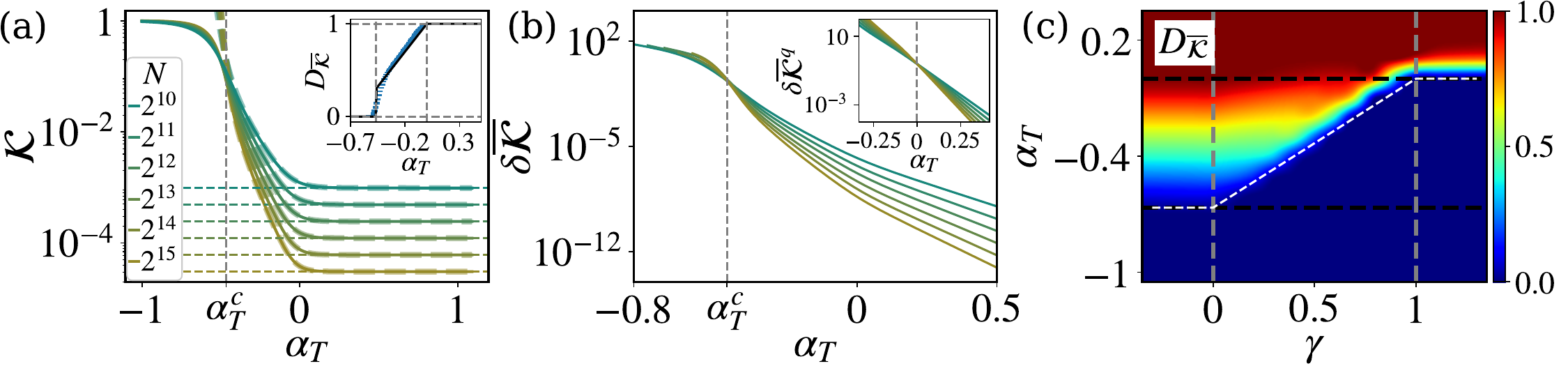}
\caption{(a)~Saturation value of the SFF shown for $\gamma = 0.4$, $\sgE = 1/2$ and different system sizes. Vertical dashed line marks $\alpha_T^c$ according to Eq.~\eqref{eq:alpha_T_c}. Horizontal lines show the infinite temperature limit $1/N$. Dashed lines indicate the annealed approximation of $\sffbar$. Inset shows the extrapolated exponent $D_{\sffbar} $ for the same $\gamma$ value as in the main panel, where vertical lines mark $\alpha_T^c$ and $\alpha_T^q$.
%===============
(b)~Relative difference $\delta{\sffbar}$ as a function of $\alpha_T$ for different system sizes. Vertical line marks $\alpha_T^c$ of Eq.~\eqref{eq:alpha_T_c}. Inset shows the difference $\delta \sffbar^q = \left|N\sffbar-1\right|$ for different system sizes and $\gamma = 0.4$ where the vertical line marks $\alpha_T^q = 0$.
%===============
(c)~Equilibrium phase diagram of the finite temperature SFF in the $\alpha_T-\gamma$ plane. Colorbar shows the different values of the scaling exponent, $D_{\overline{\mathcal{K}}}$ at different regions. The black dashed line marks $\alpha_T^q$, while the white dashed line shows the critical line of $\alpha_T^c$ corresponding to Eq.~\eqref{eq:alpha_T_c}.
}
\label{fig:sff_bar}
\end{figure}
%---------------------------
At high temperature, the annealed approximation allows us to express the saturation value as $\sffbar \approx \sffbar_\mathrm{annealed} \equiv {\mean{\nrm(T/2)}}/{\mean{\nrm(T)}^2}$.
$\sffbar$ is the IPR of $\ket{\Psi_T}$ in the energy eigenbasis, thus, quantifies the spread of the coherent Gibbs state over the energy space.
In Fig.~\ref{fig:sff_bar}(a) we show the saturation value as a function of the temperature exponent $\alpha_T$ for different system sizes and $\gamma = 0.4$ along with the annealed approximation. Figure~\ref{fig:sff_bar}(b) shows the relative difference 
$\delta{\sffbar} \equiv \left | 1- \ln\sffbar /\ln(\sffbar_\mathrm{annealed})\right|$ as a function of $\alpha_T$ for different system sizes and the $\delta{\sffbar}$ decreases (increases) with system size above (below) $\alpha_T^c$, indicating the validity of the annealed approximation beyond $\alpha_T^c$. The variation of $\delta \sffbar^q\equiv \left|N\sffbar-1\right|$ with $\alpha_T$ for different system sizes are shown in the inset of Fig.~\ref{fig:sff_bar}(b) and the finite size curves cross at a common point $\alpha_T^q$. Similar to the Gibbs weight of the ground state energy $p_1$, we define the finite size scaling exponent of the saturation value of the finite temperature SFF as
\begin{align}
    D_{\sffbar}^{(N)} \equiv -\frac{\ln \sffbar}{\ln N}
    \label{eq:D_Kbar_def}
\end{align}
and extrapolate it to thermodynamic limit to find $\lim\limits_{N\to\infty} D_{\sffbar}^{(N)} \equiv D_{\sffbar}$. The extrapolated exponent is shown in the inset of Fig.~\ref{fig:sff_bar}(a) for a fixed value of $\gamma$. We show $D_{\sffbar}$ in the $\alpha_T - \gamma$ in the plane in Fig.~\ref{fig:sff_bar}(c). It exhibits the behavior
\begin{align}
    D_{\sffbar} = \begin{cases}
        0, & \alpha_T < \alpha_T^c(\gamma) \\
        1, & \alpha_T \geq \alpha_T^q = 0,
    \end{cases}
\end{align}
where $\alpha_T^c(\gamma)$ follows the same relation with $\gamma$ as found from the scaling exponent of $p_1$. The coincidence of the two critical exponents $\alpha_T^c$ and $\alpha_T^q$ obtained from $D_{\overline{\mathcal K}}$ and $D_{p_1}$ arises because both quantities probe the concentration of the coherent Gibbs state in the energy basis. While $\langle p_1\rangle$ measures the weight of the ground state, $\overline{\mathcal K}$, being the inverse participation ratio of the CGS, tracks the effective number of levels over which the state is spread. In the frozen phase, the CGS is concentrated on the ground state, giving $p_1 \sim 1$ and $\overline{\mathcal K} \sim O(1)$, so both exponents vanish at $\alpha_T^c$. At $\alpha_T\geq\alpha_T^q=0$, the thermal weight becomes distributed over the full spectrum, with $p_n\sim N^{-1}$ and $\overline{\mathcal K}\sim N^{-1}$, so both exponents reach unity at the same temperature scale.

For \(\gamma<1\), the explicit scaling of \(p_1\) and \(\overline{\mathcal K}\) can be obtained in the regime $E_2-E_1\simeq T_c\ll T$ using the semicircular DOS given in Eq.~\eqref{eq:rho_E} in its weighted form, Eq.~\eqref{eq:rho_T_CGS_def} 
\begin{align}
    \mean{\rho_T(E)}\approx N\frac{\sqrt{4\sigma_E^2-E^2}\,e^{-E/T}}{2\pi\sgE^2 \mean{Z(T)}}\Theta(2\sigma_E-|E|)
\end{align}
one straightforwardly calculates the partition function
\begin{align}
\mean{Z(T)}\simeq N\int_{-2\sgE}^{2\sgE} \frac{\sqrt{4\sigma_E^2-E^2}}
    {2\pi\sigma_E^2} e^{-E/T} dE = \frac{N T}{\sgE} \mathrm{I}_1\left(\frac{2\sgE}{T}\right) =
    \begin{cases}
    \frac{N}{2\sqrt{\pi}\sgE^{3/2}}\,     T^{3/2}e^{2\sgE/T}, & T_c\ll T\ll T_q    \\
    N \left[1+\frac{\sgE^2}{2T^2}+O\left(\frac{\sgE^3}{T^3}\right)\right], & T\gg T_q \\
    \end{cases}
    \ , \label{eq:Z_approx}
\end{align}
where the limiting expressions for large and small $T/T_q$ are given above, $\mathrm{I}_1$ is the modified Bessel function of the first kind, and the ground state energy approximated as \(E_1\simeq-2\sigma_E\). 
This gives the normalized weighted density of states as follows
\begin{align}
    \mean{\rho_T(E)}
    = \frac{\sqrt{4\sigma_E^2-E^2}\,e^{-E/T}}{2\pi\sigma_E T\mathrm{I}_1\left(\frac{2\sigma_E}{T}\right)}\Theta(2\sigma_E-|E|).
    \label{eq:rho_T_approx}
\end{align}
Therefore using annealed approximation the average Gibbs weight of the ground state energy and the saturation value of the SFF becomes,
\begin{align}
    \mean{p_1}
    &=\frac{\mean{e^{-E_1/T}}}{\mean{Z(T)}}
    \simeq
    \begin{cases}
    \frac{2\sqrt{\pi}\sgE^{3/2}}{N T^{3/2}}, & T_c\ll T\ll T_q\\
    \frac1N\left[1+\frac{2\sgE}T+O\left(\frac{\sgE^2}{T^2}\right)\right], & T\gg T_q \\
    \end{cases}
    \\
    \overline{\mathcal K}
    &=\frac{\mean{Z(T/2)}}{\mean{Z(T)}^2}\simeq
    \begin{cases}
    \frac{\sqrt{\pi}\sgE^{3/2}}{2^{1/2}NT^{3/2}}, & T_c\ll T\ll T_q \\
    \frac1N\left[1+\frac{\sgE^2}{T^2}+O\left(\frac{\sgE^3}{T^3}\right)\right], & T\gg T_q \\
    \end{cases} ,
\end{align}
showing that the ratio $\overline{\mathcal K}/\mean{p_1}$ goes from the $2^{-3/2}$ at intermediate to the unity at high temperatures.
%For \(\gamma<1\), the explicit scaling of \(p_1\) and \(\overline{\mathcal K}\) can be obtained in the edge regime $E_2-E_1\ll T\ll 2\sigma_E$ using the semicircular DOS given in Eq.~\eqref{eq:rho_E},
%where the ground state energy is \(E_1\simeq-2\sigma_E\). 
%Near this edge, writing \(E\equiv-2\sigma_E+x\) with \(x\geq0\), the average DOS becomes
%\[
%\mean{\rho(E)}\simeq \frac{\sqrt{x}}{\pi\sigma_E^{3/2}}.
%\]
%Then using $e^{-E/T}=e^{2\sigma_E/T}e^{-x/T}$, it follows that we can approximate the average partition function as
%\begin{align}
%    \mean{Z(T)}
%    &=\mean{\sum_n e^{-\frac{E_n}{T}}}
%    \simeq N\int_{-2\sigma_E}^{2\sigma_E}
%    dE\,\mean{\rho(E)}e^{-E/T} \\
%    &\simeq \frac{N}{2\sqrt{\pi}\sigma_E^{3/2}}\,
%    T^{3/2}e^{2\sigma_E/T}.
%    \label{eq:Z(T)_T<<sgE}
%\end{align}
%Therefore using annealed approximation the average Gibbs weight of the ground state energy and the saturation value of the SFF becomes,
%\begin{align}
%    \mean{p_1}
%    &=\frac{\mean{e^{-E_1/T}}}{\mean{Z(T)}}
%    \simeq \frac{2\sqrt{\pi}\sigma_E^{3/2}}
%    {NT^{3/2}},\\
%    \overline{\mathcal K}
%    &=\frac{\mean{Z(T/2)}}{\mean{Z(T)}^2}
%    \simeq \frac{\sqrt{\pi}\sigma_E^{3/2}}
%    {2^{3/2}NT^{3/2}}
%    =2^{-5/2}\mean{p_1}.
%\end{align}

To sum up, in the parameterization \(T=2\sgE N^{\alpha_T}\), the two quantities have the equal system size scaling exponents
\begin{align}
    \boxed{D_{\overline{\mathcal K}}=D_{p_1} = 
    \begin{cases}
    0 & \alpha_T<\alpha_T^c,\\
    1-\frac{3}{2}|\alpha_T| & \alpha_T^c<\alpha_T<\alpha_T^q\equiv 0, \\
    1 & \alpha_T>\alpha_T^q\equiv 0, \\
    \end{cases}}
    \label{eq:Dp1_Dkbar_ana}
\end{align}
assuming \(\sigma_E\) is independent of \(N\). 
In addition, the above expressions explicitly show that $T_q \simeq \sgE \sim N^0$, giving the analytical prediction for $\alpha_T^q \equiv 0$.
%This explicit \(T^{-3/2}\) result applies only in the continuum edge regime. 

We show Eq.~\eqref{eq:Dp1_Dkbar_ana} along with the numerically estimated $D_{\sffbar}$ in Fig.~\ref{fig:sff_bar}(a) and $D_{p_1}$ in Fig.~\ref{fig:p1}(a), respectively, and find good agreement for $\gamma$ values close to $0$. The deviations observed for $\gamma$ values away from $0$ are attributed to finite-size effects, as the convergence to the semicircular density of states becomes increasingly slow with increasing $\gamma$.
More generally, using \(p_1=e^{-E_1/T}/Z(T)\) and \(\overline{\mathcal K}=Z(T/2)/Z(T)^2\), we can write the ratio of the two quantities as
\begin{equation}
    \frac{\overline{\mathcal K}}{p_1}
    =
    \frac{Z(T/2)e^{2E_1/T}}{Z(T)e^{E_1/T}}
    =
    \frac{1+\displaystyle\sum_{n>1}e^{-2(E_n-E_1)/T}}
         {1+\displaystyle\sum_{n>1}e^{-(E_n-E_1)/T}}.
    \label{eq:Kbar_p1_ratio}
\end{equation}
If, for the temperatures \(T\sim \Tq N^{-c}\), parametrically smaller than $\Tq$, with \(c>0\), the excited-state sum
\begin{equation}
    \sum_{n>1}e^{-(E_n-E_1)/T},
    \label{eq:excited_state_sum}
\end{equation}
remains bounded as \(N\to\infty\), the numerator in Eq.~\eqref{eq:Kbar_p1_ratio} lies between unity and its denominator. Consequently, the ratio \(\overline{\mathcal K}/p_1\) remains bounded away from zero and approaches a finite nonzero value if the excited-state sums converge. This establishes that \(\overline{\mathcal K}\) and \(p_1\) have the same system-size scaling,
\begin{equation}
    D_{\overline{\mathcal K}}=D_{p_1}.
\end{equation}
The validity of the bounded-sum assumption for the semicircular DOS and the conditions under which it could break down are discussed in Appendix~\ref{sec_appnd_ex_sum}.
%---------------------------
%   rho_T and K_dc, K_c and SFF
\begin{figure}[t]
\centering
\includegraphics[width=0.7\linewidth]{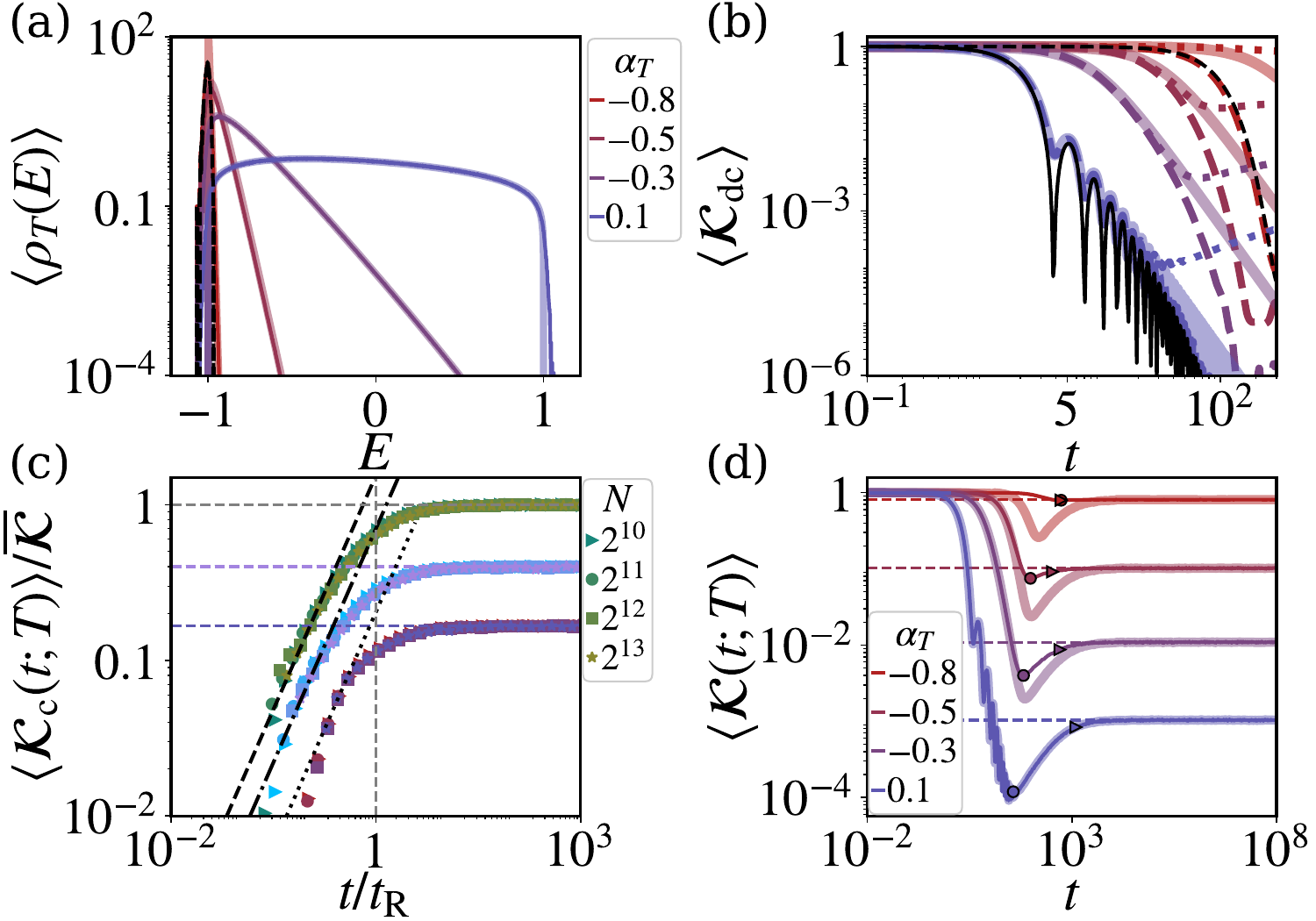}
\caption{(a)~Thermally weighted DOS for various values of $\alpha_T$, $N = 1024$, $\gamma = 0.2$ and $\sgE = 1/2$. Thin lines show numerical data whereas the thick lines show the approximation of Eq.~\eqref{eq:rho_T_approx}. The black curve shows the numerically obtained distribution of the ground state energy.
%===============
(b)~$\msffdc$ for the same parameter values as in (a). Thin lines show numerical data whereas the thick lines show the approximation of Eq.~\eqref{eq:Kdc_approx2}. Black dashed (solid) line shows the low (high) temperature limit of the disconnected part of the SFF. %\BRr
{The entire SFF are shown via dotted lines for comparison.}
%===============
(c)~$\msffc / \sffbar$ for $\gamma = 0.2$, $\alpha_T = -0.3$ and different system sizes while the time axis is scaled as $t/\tR$. The lower set of markers corresponds to $\alpha_T=-0.4$ and $\alpha_T=-0.5$ for the same system sizes where the y-axis is shifted vertically for visual clarity. The black lines indicate the reference scaling $(t/\tR)/c_\tau$, with $c_\tau\sim\mathcal{O}(1)$. The vertical line marks $t=\tR$.
%===============
(d)~disorder averaged finite temperature SFF for $\gamma = 0.2$ and system size $N = 1024$ for different values of the temperature exponent $\alpha_T$. The horizontal dashed lines mark the saturation values of the SFF at the different $\alpha_T$ values [Eq.~\eqref{eq:SFF_T_bar_1}]. Circular markers denote the dip time $\tdip$ and triangular markers denote the relaxation time $\tR$. The thick lines show the approximation of Eq.~\eqref{eq:SFF_T_parts}.
}
\label{fig:rho_T_sff}
\end{figure}
%---------------------------
%---------------------------
% Disconnected part of the SFF
\subsection{Disconnected part of the SFF}
In the \phaseA~phase ($T > \Tq$), the thermally weighted DOS of Eq.~\eqref{eq:rho_T_CGS_def} converges to the DOS. However, at lower temperature in the \phaseQ~phase ($\Tc < T < \Tq$), $\mean{\rho_T(E)}$ is positively skewed due to larger weight on the low-energy spectrum. In the \phaseF~phase ($T < \Tc$), $\mean{\rho_T(E)}$ reduces to the distribution of the ground state energy. In Fig.~\ref{fig:rho_T_sff}(a) we show $\mean{\rho_T(E)}$ for a fixed value of $\gamma$ and fixed system size with varying temperature exponent $\alpha_T$. The disconnected part can be numerically obtained from the Fourier transform of $\rho_T(E)$ as shown in Fig.~\ref{fig:rho_T_sff}(b). Thus, according to the low and high temperature behavior of $\rho_T(E)$, $\sffdc{t;T}$ can be approximated as
\begin{align}
    \mean{\sffdc{t; T}} \approx \begin{cases}
        \left |\int_{-\infty}^{\infty} dE e^{-iEt} \mathrm{P}_{E_1}(E) \right|^2, & T < \Tc \\
        \left |\int_{-\infty}^{\infty} N dE e^{-iEt} \frac{\mean{\rho(E)}e^{-E/T}}{\mean{Z(T)}} \right|^2, & T > \Tc,
    \end{cases}
    \label{eq:Kdc_approx}
\end{align}
where specifically for $T>\Tq$ it becomes 
\begin{align}
    \mean{\sffdc{t; T}} \approx \left |\int_{-\infty}^{\infty} dE e^{-iEt} \mean{\rho(E)} \right|^2.
    \label{eq:Kdc_approx_Tq}
\end{align}
In Eq.~\eqref{eq:Kdc_approx}, $\mathrm{P}_{E_1}(E)$ is the distribution of the ground state energy. The above equations are verified in Fig.~\ref{fig:rho_T_sff}(b). For \(\gamma<1\) and large \(N\), the ensemble-averaged normalized DOS follows the semicircular law given in Eq.~\eqref{eq:rho_E}, such that at temperatures $T > \Tc$, using the annealed approximation, we can separate out the averages of the numerator and the denominator and express the thermally weighted DOS as in Eq.~\eqref{eq:rho_T_approx}, with the average partition function is given by Eq.~\eqref{eq:Z_approx}.
Therefore, the disconnected part of the SFF can be calculated analytically as
\begin{align}
    \mean{\sffdc{t;T}}
    &\approx \left| \frac{1}{\mean{Z(T)}}\int_{-2\sigma_E}^{2\sigma_E}dE N\mean{\rho(E)} e^{-E/T}e^{-iEt}\right|^2\\
    &=\left|\frac{\mathrm{I}_1\!\left(2\sigma_E(T^{-1}+it)\right)}{(1+iTt)\mathrm{I}_1\left(\frac{2\sigma_E}{T}\right)}\right|^2.
    \label{eq:Kdc_approx2}
\end{align}
This analytical result is compared with the numerical Fourier transform of $\rho_T(E)$ in Fig.~\ref{fig:rho_T_sff}(b), whereas Eq.~\eqref{eq:rho_T_approx} is verified in Fig.~\ref{fig:rho_T_sff}(a). Since Eq.~\eqref{eq:rho_T_approx} is valid only for temperatures $T>\Tc$, Eq.~\eqref{eq:Kdc_approx2} agrees with the numerically estimated $\msffdc$ only in this temperature regime, while in the frozen phase, $\msffdc$ follows the low temperature approximation of Eq.~\eqref{eq:Kdc_approx}.
Figure~\ref{fig:rho_T_sff}(b) also shows that discrepancies between analytical results, Eq.~\eqref{eq:Kdc_approx2}, and numerics in the \phaseQ~phase emerge at times, longer that the dip time, defined below and, thus, they don't affect the further results.
%---------------------------
% Connected part and timescales of the SFF
\subsection{Connected part and timescales of the SFF}
The disorder averaged connected part of the SFF at finite temperature can be obtained using the relation
\begin{align}
    \mean{\sffc{t; T}} = \mean{\sff{t; T}} - \mean{\sffdc{t; T}},
\end{align}
which is responsible for the presence of the correlation hole in the time evolution of the SFF at intermediate times. In the annealed phase ($\alpha_T \geq \alpha_T^q$), the connected part of the SFF [see Eq.~\eqref{eq:SFF_T_parts}] involves the two-level form factor $b_2(\tau)$ which exhibits a linear behavior with the dimensionless relaxation time $\tauR \equiv \frac{\tR}{\tH}$~\cite{Roy2025}, for $\tau \ll \tauR$ as
\begin{align}
    b_2(\tau) = 1 - \frac{\tau}{c_\tau \tauR},
    \label{eq:b2_linear}
\end{align}
with $c_\tau$ an $\mathcal{O}(1)$ constant and $\tauR \sim N^{-\gamma} $.
In the other two phases, we assume the same linear ramp dependence and just collapse the data for different system sizes with fitted $\tauR$ scaling.
In Fig.~\ref{fig:rho_T_sff}(c) the upper set of markers show the collapse of the connected part of the SFF ($\mean{\sffc{t;T}}$) divided by the saturation value ($\sffbar$) for $\gamma = 0.2$, $\alpha_T = -0.3$ and different system sizes where the time axis is scaled as $t/\tR$. The lower set of markers corresponds to the same system sizes and $\gamma=0.2$, but with $\alpha_T=-0.4$ and $\alpha_T = -0.5$. The three sets of data exhibit the same linear behavior, indicating that the linear behavior of Eq.~\eqref{eq:b2_linear} holds even for $\alpha_T < \alpha_T^q$, only the scaling of $\tauR$ changes.

% %---------------------------
% % Timescales of the SFF
\subsection{Timescales of the finite temperature SFF}
From the behavior of the thermodynamical quantities and the scaling of the equilibrium coherent Gibbs state, we identify three distinct phases:
\begin{itemize}
    \item \phaseF~phase ($\alpha_T < \alpha_T^c$): the Gibbs weight is entirely concentrated in the ground state, $p_1=1$ and $p_{n>1}=0$, such that $\ket{\Psi_T}$ reduces to the ground state and the SFF does not evolve in time. Ensemble equivalence is absent, $\nrm^\mathrm{typ} \neq \mean{\nrm}$ [see Appendix.~\ref{sec_appnd_tydn}].

    \item \phaseQ~phase ($\alpha_T^c < \alpha_T < \alpha_T^q$): the Gibbs weights are non-uniform, $\mean{p_1}>\mean{p_2}>\dots$, while the CGS is spread over an extensive number but vanishing fraction of the energy eigenstates. Ensemble equivalence is restored, but equipartition remains absent. The Gibbs weight of the ground state exhibits a non-trivial scaling and the SFF shows peculiar temperature-dependent behavior, where its timescales depend on $\alpha_T$.

    \item \phaseA~phase ($\alpha_T>\alpha_T^q$): the Gibbs weights become approximately equal, $p_n\sim \mathcal{O}(N^{-1})$ for a finite fraction of the total number of states, and $\ket{\Psi_T}$ is homogeneously spread over the entire energy eigenbasis. Consequently, $\mean{\nrm}\approx N$ and $\mean{F}\approx -T\ln N$ and SFF shows previously reported infinite-temperature results~\cite{Roy2025}.
\end{itemize}
Thus, the thermodynamic properties and the equilibrium spreading of the coherent Gibbs state provide consistent characterizations of the same three phases.

In order to understand the overall picture better, we analyze the various timescales associated with the SFF. The spectral form factor (SFF) of a chaotic quantum system typically exhibits a dip--ramp--plateau structure~\cite{Cipolloni2023, Winer2022}. At early times, the SFF decays from its initial value, with the decay governed by the DOS. As time increases, the SFF reaches a minimum, known as the dip, after which it increases in the ramp regime due to spectral correlations, before eventually saturating to a plateau. The largest timescale of the system is the Heisenberg time, which is set by the inverse mean level spacing in the relevant energy window. To characterize these different regimes, we consider three timescales: the Zeno time $\tZ$, the dip time $\tdip$, and the relaxation time $\tR$.

The initial decay time, also known as the Zeno time, is determined via the Taylor series expansion of the SFF around $t = 0$, which reveals a universal quadratic decay~\cite{Schiulaz2019}
\begin{align}
    \mean{\sff{t; T}} = 1 - T^2\mean{\Capheat} t^2 + \mathcal{O}(t^4),\quad t\ll \tZ \equiv \frac{1}{T\sqrt{\mean{\Capheat}}}
    \label{eq:SFF_Zeno}
\end{align}
where $\Capheat$ is the heat capacity [Eq.~\eqref{eq:Cv_def}] and $\tZ$ is the Zeno time~\cite{Chiu1982}.
The Zeno time characterizes the initial short-time decay of the SFF and is determined numerically from Eq.~\eqref{eq:SFF_Zeno}. 
The dip time $\tdip$, defined as the time at which the SFF reaches its minimum, marks the intersection of the initial decay and the ramp regime. 
Finally, the relaxation time $\tR$, defined in Eq.~\eqref{eq:b2_linear}, characterizes the onset of the plateau, where the SFF (or its connected part) approaches its saturation value. In systems exhibiting a correlation hole, these timescales distinguish the initial decay, the development of spectral correlations, and the eventual saturation of the SFF. 

In Fig.~\ref{fig:rho_T_sff}(d) we show the time evolution of the disorder averaged SFF, both disconnected and connected parts, for a fixed system size and fixed value of $\gamma$ in the different temperature regimes. In this figure we also show the numerically obtained relaxation times ($\tR$) and dip times ($\tdip$) via markers. At sufficiently high temperature, the disorder averaged SFF has good agreement with Eq.~\eqref{eq:SFF_T_parts}, as shown in the same figure. Here $b_2(\tau)$ in Eq.~\eqref{eq:SFF_T_parts} is estimated numerically from the Fourier transform of the two-level cluster function~\cite{Roy2025, Mehta2004book}. However, at low temperature, the SFF is dominated by the ground and low-lying excited states. Such low-energy states have a weaker energy correlation compared to the bulk states, leading to a shallower correlation hole [shown in Fig.~\ref{fig:rho_T_sff}(d)]. Consequently, at low temperature, Eq.~\eqref{eq:SFF_T_parts} correctly predicts the initial decay and equilibration but not the correlation hole.

\begin{figure}[t]
\centering
\includegraphics[width=0.7\linewidth]{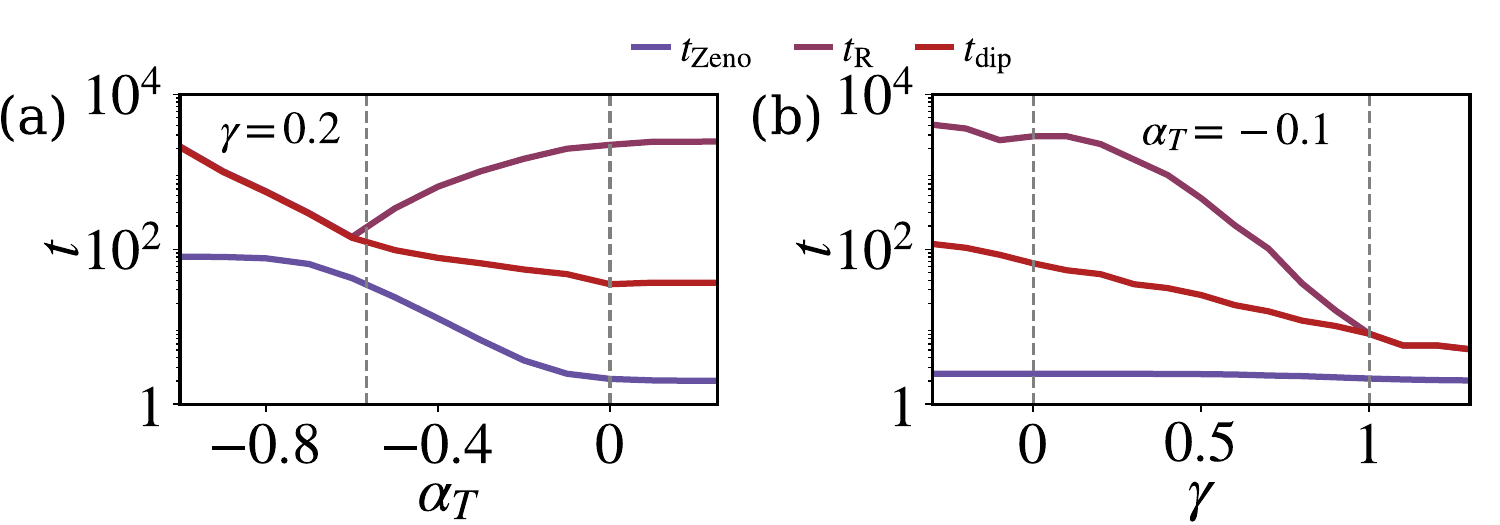}
\caption{Finite temperature SFF timescales for $\sgE = 1/2$. (a)~Different timescales as a function of $\alpha_T$ for $\gamma = 0.2$ and system size $N = 1024$. Vertical gray dashed lines indicate $\alpha_T^c$ [Eq.~\eqref{eq:alpha_T_c}] and $\alpha_T^q$.
%===============
(b)~Same timescales as in (a) as a function of $\gamma$ for $\alpha_T = -0.1$ and $N = 1024$. Vertical gray dashed lines mark the range of the NEE regime.
%\imk{In both panels, shall we add the predictions from the Table I? At the same time it will be useful to add the phase boundaries to both plots. Otherwise the plots are quite unclear and even useless for a general reader.}
%\BR{Since we do not know the temperature-dependent prefactors of the timescales, we cannot make a reliable quantitative comparison with the predictions in Table I.}
}
\label{fig:sff_timescales}
\end{figure}

\begin{figure}[t]
\centering
\includegraphics[width=\linewidth]{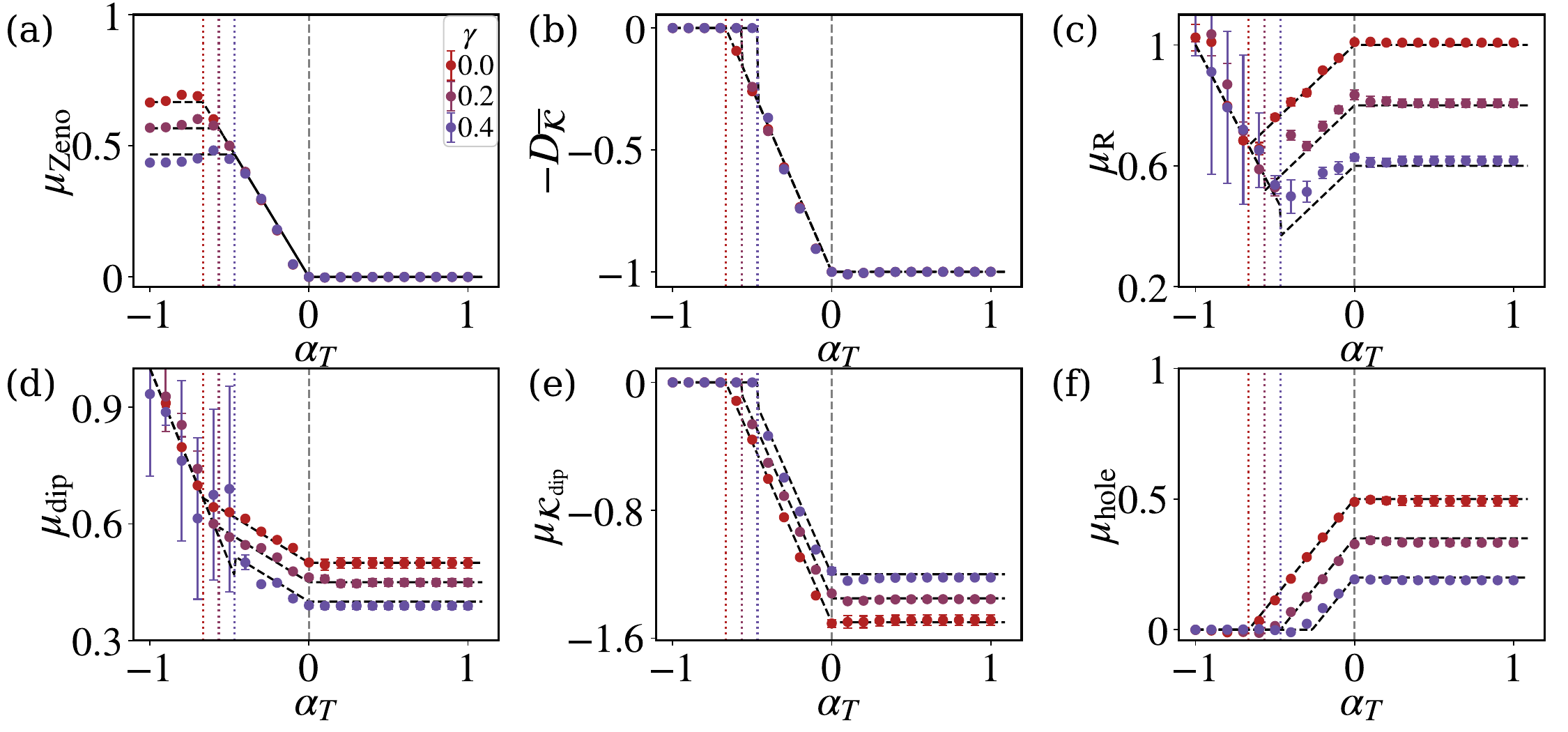}
\caption{System-size scaling exponents of the characteristic SFF timescales and the relative depth of the correlation hole. (a)~Scaling exponent of the Zeno time as a function of $\alpha_T$ for various values of $\gamma$. Vertical dotted lines mark $\alpha_T^c$ of the corresponding $\gamma$ given in Eq.~\eqref{eq:alpha_T_c}.
%===============
(b), (c), (d), (e) and (f)~Scaling exponent of $\sffbar$, $\tR$, $\tdip$, $\mean{\mathcal{K}(\tdip)}$,  and the relative depth of the correlation hole [Eq.~\eqref{eq:corr_depth}] as a function of $\alpha_T$ for the same values of $\gamma$ as in (a). In all panels the black lines show the proposed forms of the exponents as given in Table~\ref{tab:sff_timescales}. Specifically $D_{\sffbar}$ is known analytically from Eq.~\eqref{eq:Dp1_Dkbar_ana}, $\mu_\mathrm{Zeno}$ and $\mu_\mathrm{R}$ are numerically obtained, whereas $\mu_\mathrm{dip}$, $\mu_{\mathcal{K}_\mathrm{dip}}$ and $\mu_\mathrm{hole}$ are analytically known from Eqs.~\eqref{eq:t_dip_ana}, \eqref{eq:K(t_dip)_ana} and \eqref{eq:d_hole_pred} using the values of $D_{\sffbar}$ and $\mu_\mathrm{R}$.
%\imk{I am sorry: now this comment is in the right figure caption. Please address it: I suggest to order the panels in the same way as the Table columns: from what we know analytically (Zeno, $\sffbar$ to what we use ($\tR$) to extract the last scalings ($t_{\rm dip}$, $\mathcal{K}_\mathrm{dip}$, and $d_{\rm hole}$). What do you think?}
}
\label{fig:sff_timescales_exponents}
\end{figure}

Figure~\ref{fig:sff_timescales}(a) shows the characteristic timescales of the SFF, namely, the Zeno, dip, and relaxation times, as a function of $\alpha_T$ for a fixed value of $\gamma$. We observe that all three timescales become independent of $\alpha_T$ for $\alpha_T>\alpha_T^q$. 
%Furthermore, for $\alpha_T<\alpha_T^c$, the dip and relaxation times coincide, indicating the disappearance of a distinct correlation hole. 
Figure~\ref{fig:sff_timescales}(b) shows the same timescales as a function of $\gamma$ for a fixed value of $\alpha_T$. 
Furthermore, we observe that the dip and relaxation times coincide for $\alpha_T<\alpha_T^c$ in~(a) and for $\gamma\geq 1$ in~(b), indicating the disappearance of a distinct correlation hole. In Fig.~\ref{fig:sff_timescales_exponents}, we show the system-size scaling exponents of the characteristic timescales and related quantities as a function of $\alpha_T$ for various values of $\gamma$. Panels (a), (c) and (d) show the exponents of the Zeno time, the relaxation time and the dip time, respectively, whereas panels (b) and (e) show the exponents of the saturation value, $\sffbar\sim N^{-D_{\sffbar}}$, and of the SFF at the dip, $\mean{\mathcal{K}(\tdip)}\sim N^{\mu_{\mathcal{K}_\mathrm{dip}}}$, respectively. We also quantify the relative depth of the correlation hole as a measure of the strength of spectral correlations. Therefore, a unitless dynamical measure of the energy correlation is
\begin{align}
	d_\mathrm{hole} = \frac{\sffbar}{\mean{\sff{\tdip}}} \sim N^{\mu_{\rm hole}} \ , \quad \mu_{\rm hole}=-\mu_{\mathcal{K}_\mathrm{dip}}-D_{\sffbar}
	\label{eq:corr_depth}
\end{align}

%=======================================
\begin{table}[b!]
\centering
\caption{Numerically obtained system-size scaling exponents of the characteristic SFF timescales, saturation value of the SFF and the relative depth of the correlation hole. Here, $\tZ\sim N^{\mu_{\rm Zeno}}$, Eq.~\eqref{eq:SFF_Zeno}, $\sffbar\sim N^{-D_{\sffbar}}$, Eq.~\eqref{eq:Dp1_Dkbar_ana}, $\tR\sim N^{\mu_{\rm R}}$, Eq.~\eqref{eq:b2_linear}, $\tdip\sim N^{\mu_{\rm dip}}$, Eq.~\eqref{eq:t_dip_ana}, and $\mean{\mathcal{K}(\tdip)}\sim N^{\mu_{\mathcal{K}_\mathrm{dip}}}$, Eq.~\eqref{eq:K(t_dip)_ana}, while $\mu_{\rm hole}$, Eq.~\eqref{eq:d_hole_pred}, characterizes the system-size scaling of the relative depth of the correlation hole. The exponent $\alpha_T^c(\gamma)$ is given in Eq.~\eqref{eq:alpha_T_c}.
}
\label{tab:sff_timescales}
\resizebox{\textwidth}{!}{%
\begingroup
\renewcommand{\arraystretch}{1.5}
\begin{tabular}{c|c|c|c|c|c|c}
\hline
$\alpha_T$ 
& $\mu_{\rm Zeno}$ 
& $D_{\sffbar}$ 
& $\mu_{\rm R}$ 
& $\mu_{\rm dip}$ 
& $\mu_{\mathcal{K}_\mathrm{dip}}$ 
& $\mu_{\rm hole}$ \\
\hline
$\alpha_T < \alpha_T^c(\gamma)$
&
$-\alpha_T^c(\gamma)$
&
$0$
&
$-\alpha_T$
&
$-\alpha_T$
&
$0$
&
$0$
\\
\hline
$\alpha_T^c(\gamma)\leq\alpha_T<0$
&
$-\alpha_T$
&
$1+\frac{3\alpha_T}{2}$
&
$1-\gamma+\frac{\alpha_T}{2}$
&
$\frac{1}{2}-\frac{\gamma}{4}-\frac{\alpha_T}{4}$
&
$-\frac{3}{2}+\frac{3\gamma}{4}-\frac{9\alpha_T}{4}$
&
$\frac12-\frac{3\gamma}{4}+\frac{3\alpha_T}{4}$
\\
\hline
$\alpha_T\geq 0$
&
$0$
&
$1$
&
$1-\gamma$
&
$\frac{1}{2}-\frac{\gamma}{4}$
&
$-\frac{3}{2}+\frac{3\gamma}{4}$
&
$\frac12-\frac{3\gamma}{4}$
\\
\hline
\end{tabular}
\endgroup
}
\end{table}
%---------------------------

%---------------------------
%   SFF timescales
%---------------------------
%   SFF timescales exponents
%---------------------------

In Fig.~\ref{fig:sff_timescales_exponents}(f) we show the variation of the scaling exponent of $\mu_{\rm hole}$ %$d_\mathrm{hole}\sim N^{\mu_{\rm hole}}$, 
depending on $\alpha_T$. Numerically observed behavior of the scaling exponents are given in Table~\ref{tab:sff_timescales}. For $\alpha_T>\alpha_T^q=0$, all exponents attain their infinite-temperature values~\cite{Roy2025, Note}. At infinite temperature, the numerical results for the scaling exponents agree with our predicted forms for small values of $\gamma$. Thus, here in the figures we show the data only for $\gamma \lesssim 0.5$. 
Now using the approximation of Eq.~\eqref{eq:Kdc_approx2} the initial envelope of the SFF follows a power law decay $\sim (\min(T,2\sgE)t)^{-3}$ and the two-level form factor has a linear growth at small time given by Eq.~\eqref{eq:b2_linear}. Thus, focusing on $T\ll 2\sgE$ ($\alpha_T<\alpha_T^q = 0$), from the intersection of the slope ($\sim (Tt)^{-3}$) and the ramp, we can estimate the dip time as $\tdip \sim N^{\mu_{\rm dip}}$

\begin{align}
    (T\tdip)^{-3} \sim \left [1-b_2\left(\frac{\tdip}{\tH}\right)\right ]\sffbar = \frac{\tdip\sffbar}{\tR} \Rightarrow 
    \boxed{\mu_{\rm dip} = \frac{1}{4}\left(\mu_\mathrm{R} + D_{\sffbar} - 3\alpha_T\right)}= 
    \begin{cases}
    \frac{1}{4}\left(\mu_\mathrm{R} - 3\alpha_T\right), & \alpha_T<\alpha_T^c, \\
    \frac{1}{4}\left(\mu_\mathrm{R} + 1 - \frac{3}{2}\alpha_T\right), & \alpha_T>\alpha_T^c,
    \end{cases}
    %\tdip \sim N^{\frac{\mu_\mathrm{R}}{4} + \frac{D_{\sffbar}}{4} - \frac{3\alpha_T}{4}},
    \label{eq:t_dip_ana}
\end{align}
At the dip time the SFF has a value 
\begin{align}
\mean{\sff{\tdip}} \sim \frac{\tdip\sffbar}{\tR} \equiv N^{\mu_{\mathcal{K}_\mathrm{dip}}}
\Rightarrow \boxed{\mu_{\mathcal{K}_\mathrm{dip}}=-\frac{3}{4}\left(\mu_\mathrm{R} + D_{\sffbar} +\alpha_T\right)} = 
    \begin{cases}
     -\frac{3}{4}\left(\mu_\mathrm{R} + \alpha_T\right), & \alpha_T<\alpha_T^c, \\
     -\frac{3}{4}\left(\mu_\mathrm{R} + 1 + \frac{5}{2}\alpha_T\right),& \alpha_T>\alpha_T^c.
    \end{cases}
\label{eq:K(t_dip)_ana}
\end{align}
Here we used the prediction for $D_{\sffbar}$, Eq.~\eqref{eq:Dp1_Dkbar_ana}.
Then, from Eq.~\eqref{eq:corr_depth}, the relative depth of the correlation hole for $\alpha_T \leq 0$ scales as,
\begin{align}
    d_\mathrm{hole}\equiv N^{\mu_{\rm hole}}  \Rightarrow 
    \boxed{\mu_{\rm hole} =\frac{1}{4}\left(3\mu_\mathrm{R}+ 3\alpha_T-D_{\sffbar}\right)} = 
    \begin{cases}
    \frac{3}{4}\left(\mu_\mathrm{R}+\alpha_T\right), & \alpha_T<\alpha_T^c, \\
    \frac{1}{4}\left(3\mu_\mathrm{R}+ \frac{3}{2}\alpha_T -1\right), & \alpha_T>\alpha_T^c,
    \end{cases}
    %\sim N^{-\frac{1}{4}D_{\sffbar} + \frac{3}{4} \mu_\mathrm{R}+ \frac{3}{4} \alpha_T},
    \label{eq:d_hole_pred}
\end{align}

Finally, the only missing piece is the scaling of $\tR \sim N^{\mu_{\rm R}}$ which numerical values we take from Table~\ref{tab:sff_timescales} to obtain
the expressions, consistent with those that has been extracted numerically in Table.~\ref{tab:sff_timescales}. 
%\begin{align}
%    \mu_\mathrm{dip} \sim 
%    \begin{cases}
%        \frac{1}{2} - \frac{\alpha_T}{4} - \frac{\gamma}{4} & \alpha_T < 0 \\
%        \frac{1}{2} - \frac{\gamma}{4} & \alpha_T \geq 0.
%    \end{cases}
%    \label{eq:mu_dip}
%\end{align}
%Then, the relative depth of the correlation hole for $\alpha_T \geq 0$ scales as,
%\begin{align}
%    d_\mathrm{hole}\imkr{\equiv N^{\mu_{\rm hole}}  \Rightarrow 
%    \boxed{\mu_{\rm hole} =-\frac{1}{4}D_{\sffbar} + \frac{3}{4} \mu_\mathrm{R}+ \frac{3}{4} \alpha_T },}
%    %\sim N^{-\frac{1}{4}D_{\sffbar} + \frac{3}{4} \mu_\mathrm{R}+ \frac{3}{4} \alpha_T},
%    \label{eq:d_hole_pred}
%\end{align}
%therefore, the exponent of $d_\mathrm{hole}$ goes as 
%\begin{align}
%    \mu_\mathrm{hole} \sim
%    \begin{cases}
%        -\frac{3\gamma}{4} +\frac12 + \frac{3\alpha_T}{4} & \alpha_T^c(\gamma)\leq\alpha_T<0\\
%         \frac12-\frac{3\gamma}{4} & \alpha_T \geq 0.
%    \end{cases}
%    \label{eq:mu_hole}
%\end{align}
%Thus Eq.~\eqref{eq:mu_dip} and Eq.~\eqref{eq:mu_hole} are in agreement with Table.~\ref{tab:sff_timescales}. 
%---------------------------
% Absence of mobility edge
\subsection{Absence of mobility edge}
Several quantities can be employed to investigate the absence of a ME that separates ergodic states from the localized states. For instance, the energy-resolved IPR and the fractal dimension ($D_2$) provides strong information about spectral homogeneity of the eigenstate properties in the NEE regime of the \bte. On the other hand, the energy resolved averaged level-spacing ratio $\mean{r}$ that characterizes short-range spectral correlations, also shows no evidence of a ME~\cite{Das2023}. 
In this work, we compare the energy dependence of the IPR~\cite{Das2023} with the temperature dependence of the relative depth of the correlation hole~\cite{Roy2026}. This comparison tests whether the absence of a mobility edge indicated by the energy-resolved IPR is also reflected in the finite-temperature SFF. %\BRr
{Accordingly we take the system-size scaling exponent of $d_\mathrm{hole}$
and normalize it by its value at $\gamma=0$ and $T\to\infty$:
\begin{align}
    \overline{\mu}_\mathrm{hole}
    = \frac{\mu_\mathrm{hole}(\gamma,T)}
    {\mu_\mathrm{hole}(\gamma=0,T\rightarrow\infty)}
    \in [0,1].
\end{align}
Values of $\overline{\mu}_\mathrm{hole}$ near zero indicate an absent or shallow correlation hole, while values of order unity indicate a pronounced hole.}

%\BRr
{In Ref.~\onlinecite{Das2023}, it was shown that localized and extended states coexist within an energy window $(-E_G',E_G')$ centered around $E=0$. We use the energy-temperature duality to determine the temperature $T_G$ (or equivalently, the temperature exponent $\alpha_T^G$) at which the ensemble-averaged thermal energy reaches the boundary of the energy window $|E|<E_G'$, i.e.,
\begin{align}
    -E_G'=\left.\overline{\langle E\rangle_T}\right|_{T=T_G}.
    \label{eq:mean_E}
\end{align}
Within the annealed approximation, we use the ensemble-averaged thermally weighted DOS, Eq.~\eqref{eq:rho_T_approx}, to estimate the ensemble-averaged thermal energy at $T\gg\sigma_E$ (see Appendix~\ref{sec_appnd_meanE} for details):
\begin{align}
    \overline{\langle E\rangle_T}
    \simeq -\frac{\sigma_E^2}{T}
    +\mathcal{O}\left(\frac{\sigma_E^4}{T^3}\right).
    \label{eq:<E>=sgE^2/T}
\end{align}
Thus, for $\gamma<1$, using $E_G'\sim N^{-(1-\gamma)/2}$ and the temperature scaling $T=2\sgE N^{\alpha_T}$, we find
\begin{align}
\alpha_T^G=\frac{1-\gamma}{2}>0,
\qquad \gamma<1.
\label{eq:alpha_T^G}
\end{align}}

\begin{figure}[t]
\centering
\includegraphics[width=0.7\linewidth]{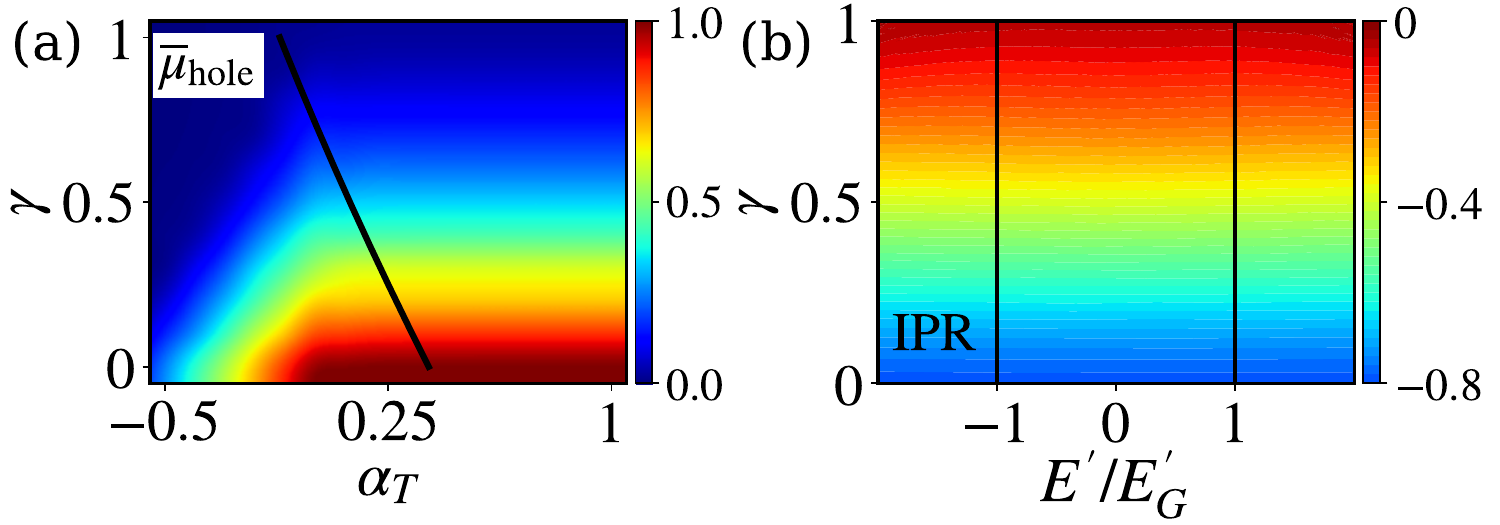}
\caption{(a)~Color-map of $\overline{\mu}_\mathrm{hole}$ in the $\gamma - \alpha_T$ plane for $N = 32768$. The black line marks $\alpha_T^G$.
%=====\gamma - \alpha_T= plane= The black line marks $\alpha_T^G$.========
(b)~Color map of the energy-resolved quantity $\mean{\ln(\mathrm{IPR})}/\ln N$
in the $\gamma-E'$ plane for $N = 8192$, where the x-axis is scaled as $E' / E_G'$. The black line indicates $(-E_G', E_G')$, the energy band of localized states.
%\imk{It seems that the figure is again the old one with invisible results in panel (b). Please double check.}
}
\label{fig:dhole_ipr}
\end{figure}

%\BRr
{In Fig.~\ref{fig:dhole_ipr}(a), we show the normalized relative depth of the correlation hole in the $\gamma-\alpha_T$ plane. The solid black line in Fig.~\ref{fig:dhole_ipr}(a) indicates $\alpha_T^G$. Here, the positive energy boundary $E_G'$ corresponds to $-T_G$, invoking the concept of negative temperatures in systems with bounded spectra~\cite{ramsey1956, mondragon2015, abraham2017}.}

%\BRr
{At fixed $\gamma$, a mobility edge within the energy window $(-E_G',E_G')$ would be expected to manifest as a change in $\overline{\mu}_{\rm hole}$ from $0$ to $1$ across the corresponding temperature regime, i.e., to the right of the black line. However, our analysis shows that $\overline{\mu}_{\rm hole}$ remains nearly constant for $\alpha_T>\alpha_T^q=0$. Since $\alpha_T^G>\alpha_T^q$ for $\gamma<1$, the region to the right of the black line lies entirely in this temperature-independent regime. Thus, $\overline{\mu}_{\rm hole}$ exhibits no crossover from $0$ to $1$ for $T>T_G$, indicating that the coexistence of localized and extended states within $(-E_G',E_G')$ does not give rise to a mobility-edge signature in the thermally averaged correlation-hole depth.
}

Figure~\ref{fig:dhole_ipr}(b) shows the energy-resolved IPR in the $\gamma$-$E'$ plane, defined as
\begin{align}
    I(E) = \frac{1}{N\rho(E)}\sum_{j=1}^{N}I_j\delta(E-E_j),
\end{align}
with energy $E' / E_G'$ on the horizontal axis and $\gamma$ on the vertical axis.
The black curves mark the boundaries $E'=\pm E_G'$ of the energy window discussed in Ref.~\onlinecite{Das2023}, where localized and extended states coexist. Within this window, the IPR shows no energy-dependent crossover from small to order-one values.

The comparison shows that the IPR has no such crossover within the marked energy window, while $\overline{d}_\mathrm{hole}$ shows no signature of mobility edge within the corresponding temperature boundary and remains nearly constant across the associated range. Thus, the finite-temperature SFF reflects the absence of a mobility edge indicated by the energy-resolved IPR in the NEE regime of the \bte.
Though, here one should note that, as soon as $T_G\gg T_q \simeq \sgE$, Eq.~\eqref{eq:alpha_T^G}, at those temperatures the SFF shows only infinite-temperature results and cannot be sensitive to any mobility edge. Nevertheless, as we show in the next section, even in the absence of the mobility edge, the finite-temperature SFF can give us the information about the infinite-temperature spectral properties with much less computation time.
% dhole IPR
%---------------------------
%---------------------------
%=============================
\section{Finite temperature Lanczos method}
\label{sec:lanczos}
\begin{figure}[t]
\centering
\includegraphics[width=0.7\linewidth]{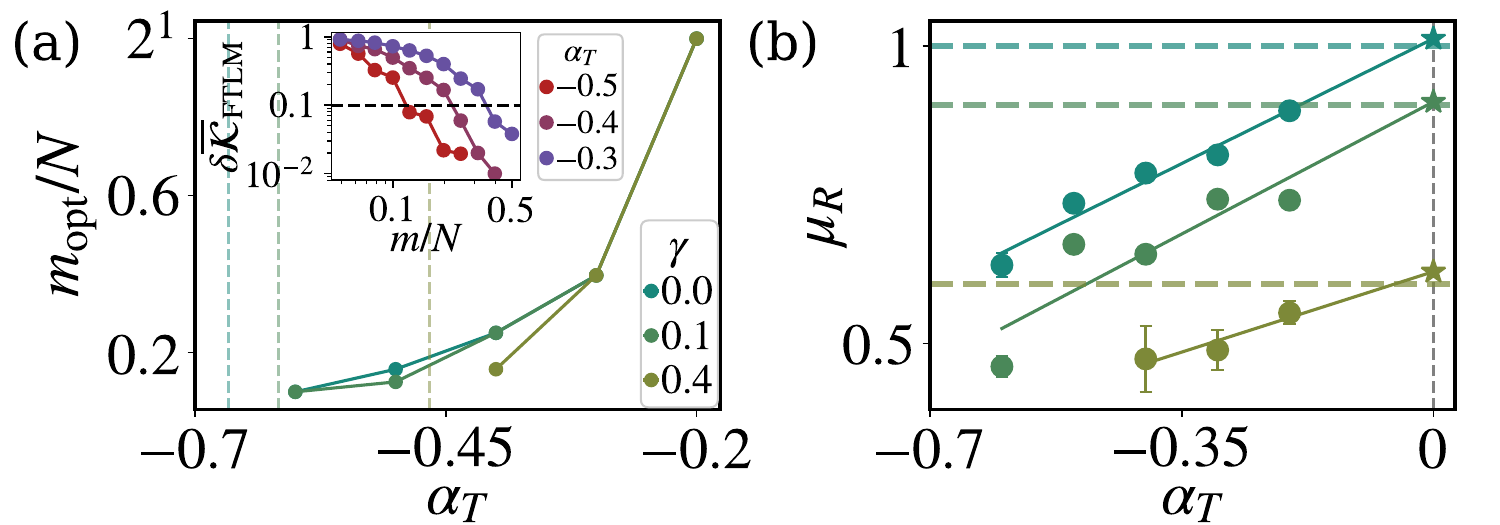}
\caption{(a)~The optimized value of $m$, divided by the system size $N$ as a function of $\alpha_T$ for $N = 1024$, $\sgE = 1/2$  and different values of $\gamma$. Vertical dashed lines mark $\alpha_T^c$ given in Eq.~\eqref{eq:alpha_T_c}. Inset shows $\delta \sffbar_\mathrm{FTLM}$ as a function of $m/N$ for different values of $\alpha_T$ and $\gamma = 0.2$ with system size $N = 1024$. The horizontal dashed line marks the threshold which we set to be $0.1$.
%===============
(b)~The extrapolation $\mu_R$ to $\alpha_T = \alpha_T^q$ for the same values of $\gamma$ used in panel (a). The markers at $\alpha_T=\alpha_T^q$ indicate the extrapolated values. Horizontal lines mark $1-\gamma$.
}
\label{fig:FTLM}
\end{figure}

%=============================
The finite-temperature SFF, defined in Eq.~\eqref{eq:SFF_T_def} through the analytic continuation of the partition function $Z(T,it)$, requires evaluating the trace $\mathrm{Tr}\left[e^{-(\frac{1}{T}+it)H}\right]$, which becomes computationally expensive for large matrix sizes. To overcome this, we use the finite-temperature Lanczos method (FTLM)~\cite{Jaklic1994}, where the full trace is replaced by an average over $N_v$ random vectors $\{\ket{r}\}$. For each vector, $m$ Lanczos iterations generate a tridiagonal matrix whose eigenvalues $\{E_l^{(r)}\}$ and weights $\{|\langle \phi_l^{(r)}|r\rangle|^2\}$ approximate the spectrum of the full Hamiltonian, so that
\begin{equation}
	Z(T, it)
	\simeq
	\frac{1}{N_v}
	\sum_{r=1}^{N_v}
	\sum_{l=1}^{m}
	|\langle \phi_l^{(r)}|r\rangle|^2
	e^{-\left(\frac{1}{T}+it\right)E_l^{(r)}},
    \label{eq:ZT_FTLM}
\end{equation}
from which the SFF follows via Eq.~\eqref{eq:SFF_T_def}. The minimum number of iterations needed to reproduce the exact-diagonalization SFF within a $10\%$ relative error is denoted by $m_\mathrm{opt}$ [see Appendix~\ref{sec_appnd_FTLM} for details]. As shown in Fig.~\ref{fig:FTLM}(a), $m_\mathrm{opt}$ grows with $\alpha_T$ and approaches $\mathcal{O}(N)$ as the Gibbs distribution becomes uniform and probes a larger fraction of the spectrum. However for $\alpha_T \le \alpha_T^q$ we show that $m_\mathrm{opt}\ll N$, i.e. the number of Lanczos steps required to faithfully reproduce the finite temperature SFF is much smaller than the system size. In the inset we show the relative error $\delta \sffbar_\mathrm{FTLM}$ [defined in Appendix~\ref{sec_appnd_FTLM}] as a function of number of iterations divided by the system size, $m/N$ for different values of $\alpha_T$ where the horizontal line marks the threshold, illustrating the increase in $m_\mathrm{opt}$ with increasing $\alpha_T$.

With this Lanczos method, we can extract the infinite-temperature timescale exponents from the finite-temperature counterparts without diagonalizing the full $N\times N$ matrices. As shown in Fig.~\ref{fig:sff_timescales_exponents}(b), the relaxation-time exponent $\mu_\mathrm{R}$, defined through $\tR\sim N^{\mu_\mathrm{R}}$, varies linearly with $\alpha_T$ between $\alpha_T^c$ and $\alpha_T^q$. We therefore compute $\tR$ with FTLM in this window, where only $m_\mathrm{opt}\ll N$ iterations are needed, and extrapolate $\mu_\mathrm{R}$ linearly to $\alpha_T=0$ [Fig.~\ref{fig:FTLM}(b)]. The extrapolated value agrees well with the infinite-temperature prediction $\mu_\mathrm{R}=1-\gamma$, so the infinite-temperature timescales can be estimated for system sizes far beyond the reach of exact diagonalization.
%---------------------------
% FTLM
%---------------------------
%=============================
\section{Conclusion}
\label{sec:conclusion}
%=============================
In this work, we look at the thermodynamics and the finite-temperature spectral form factor (SFF) of the $\beta$ ensemble, where we scale the temperature as $T\sim N^{\alpha_T}$. By studying the system-size scaling of the ground-state Gibbs weight, $p_1 \sim N^{-D_{p_1}}$, we identify three thermal regimes: a frozen phase, a quenched phase, and an annealed phase. In the frozen phase ($\alpha_T < \alpha_T^c$), the Gibbs weight of the ground state dominates the thermodynamics of the system, and $p_1 \sim 1$, where neither ensemble equivalence nor equipartition of Gibbs weights holds. In the quenched phase ($\alpha_T^c < \alpha_T < \alpha_T^q$), the coherent Gibbs state spreads over an extensive but vanishing fraction of the eigenstates, and ensemble equivalence is restored while equipartition is still absent. In the annealed phase ($\alpha_T > \alpha_T^q$), the Gibbs weights become approximately uniform, i.e., $p_n \sim \mathcal{O}(1/N)$, for a finite fraction of the total number of states. The transition from the frozen to the quenched phase is governed by the temperature exponent $\alpha_T^c$ which is a function of the system parameter $\gamma$, whereas the exponent $\alpha_T^q$ which separates the quenched from the annealed phase is independent of $\gamma$ for a fixed spectral bandwidth. The freezing transition is further reflected in a nonanalyticity of the free energy, a maximum of the heat capacity that grows with $N$ for $\gamma>0$, and a loss of self-averaging of the partition function.

We further establish how these thermal regimes are manifested in the finite-temperature SFF, whose saturation value is the IPR of the coherent Gibbs state in the energy eigenbasis, scaling as $\sim N^{-D_{\sffbar}}$. The ground state Gibbs weight and the equilibrium SFF both probe the same spreading of the state over the spectrum and hence give rise to the same system size scaling exponents $D_{p_1} = D_{\sffbar}$, as a function of the temperature exponent $\alpha_T$ and we derived the analytical form of this scaling.

The finite-temperature SFF exhibits distinct dynamical behavior across the three thermal regimes. In particular, in the frozen phase, the SFF is dominated by the ground state, and consequently the SFF exhibits no evolution in time and remains unity throughout. We further derive analytical scaling forms for the dip time and the relative depth of the correlation hole, relating the scaling exponents of these quantities to the temperature exponent $\alpha_T$, the system parameter $\gamma$ and the relaxation time exponent $\mu_\mathrm{R}$. These relations provide a consistent picture of the system size scaling of the characteristic features of the SFF across the various temperature regimes, summarized in Table~\ref{tab:sff_timescales}. We also look at the temperature dependence of the correlation hole in relation to the energy-resolved IPR and find that there is no signature of a mobility edge in the energy window where both localized and extended states are present. This is in agreement with the previously established absence of a mobility edge in the NEE regime of the $\beta$ ensemble.

Finally, we demonstrate the application of the finite-temperature Lanczos method to calculate the finite-temperature SFF without requiring the full energy spectrum. In particular, at low temperatures, the number of Lanczos iterations required to reproduce the SFF is much smaller than the system size, making this approach computationally less expensive than exact diagonalization. More importantly, we show that the temperature dependence of the relaxation-time exponent in the low-temperature regime can be used to extrapolate its value to the infinite-temperature limit, $\alpha_T=0$. The extrapolated exponent agrees well with the infinite-temperature prediction $\mu_\mathrm{R}=1-\gamma$. Thus, even in the absence of a mobility edge, the finite-temperature SFF, together with finite-temperature Lanczos method, provides an efficient way to estimate infinite-temperature spectral properties without diagonalizing the full Hamiltonian. This is particularly useful for studying the ratio $t_\mathrm{R}/t_\mathrm{H}$, which serves as an indicator of the Anderson transition in systems with anomalous energy correlations~\cite{Roy2025}.

%=============================
\section{Acknowledgments}
%=============================
I.~M.~K.~acknowledges support by the French National Research Agency (ANR) under the France 2030 program, in subprograms ``Choose France for Science", reference ``ANR-25-CFFS-0019'', and the TIRIS (Toulouse Initiative for Research's Impact on Society) Blockbuster Programme, reference ``ANR-22-EXES-0015'', the Occitanie region, the European Regional Development Fund (ERDF), and the European Union NextGenerationEU. A.~K.~D.~acknowledges support from the Leverhulme Trust Research Project Grant RPG-2025-063.

%=============================
\appendix
%===========================================================
\section{Condition for equipartition}\label{sec_appnd_p1_1/N}
%===========================================================
The energy levels are ordered as $E_1\leq E_2\leq\cdots\leq E_N$, which implies $p_1\geq p_2\geq\cdots\geq p_N$. Let $p_m$ be parametrically smaller than the equipartition weight $1/N$ for some $1\leq m\leq N$, i.e.,
\begin{align}
    p_m=c_m N^{-\alpha_m},
    \qquad
    c_m\sim\mathcal{O}(1),
    \qquad
    \alpha_m>1.
\end{align}
Due to the monotonicity of the Gibbs weights, all subsequent weights $p_{m+1},p_{m+2},\ldots,p_N$ are also parametrically smaller than $1/N$. Consequently,
\begin{align}
    \sum_{j=m}^{N}p_j
    \leq (N-m+1)p_m
    =\mathcal{O}(N^{1-\alpha_m})
    \longrightarrow 0
\end{align}
in the thermodynamic limit. Therefore, the normalization condition $\sum_{j=1}^{N}p_j=1$ requires
\begin{align}
    \sum_{j=1}^{m-1}p_j=1-\mathcal{O}(N^{1-\alpha_m}).
\end{align}
Since $p_j\leq p_1$ for all $j$, we obtain
\begin{align}
    (m-1)p_1\geq 1-\mathcal{O}(N^{1-\alpha_m}).
\end{align}
Thus, if
\begin{align}
    p_1\propto N^{-1}
    \qquad
    (\text{i.e., }D_{p_1}=1),
\end{align}
the number of states $m$ whose Gibbs weights are not parametrically smaller than $1/N$ must constitute a finite fraction $m \simeq O(N)$ of the total number of states in the thermodynamic limit.

%===========================================================
\section{Thermodynamic quantities}\label{sec_appnd_tydn}
%===========================================================
The entropy is fundamentally defined from the number of accessible microstates in a given realization. If $\mathcal{N}(E)dE$ denotes the number of states in an energy window $[E,E+dE]$ with $dE \ll 2\sgE$, the disorder averaged microcanonical entropy is
\begin{align}
    S(E)=\mean{\ln \left(\mathcal{N}(E)dE\right)}.
\end{align}
Since $\mathcal{N}(E)dE=N\rho(E)dE$ is determined by the DOS $\rho(E)$, defined in the main text, this gives,
\begin{align}
    S(E)=\ln N+\mean{\ln\rho(E)}+\ln dE.
\end{align}
Then assuming that the DOS is self-averaging in the thermodynamic limit, we have
\begin{align}
    \mean{\ln\rho(E)}
    \simeq
    \ln\mean{\rho(E)}.
\end{align}
Thus by introducing the energy scale $\EM$ through $\mean{\rho(-\EM)}=(N dE)^{-1}$, the ensemble-averaged microcanonical entropy becomes~\cite{Derrida1980, Derrida1981, Roy2026}
\begin{align}
    \SMC(E) & = \begin{cases}
        0, & E \leq -\EM\\
        \ln N + \ln \mean{\rho(E)} + \ln dE, & E > -\EM.
    \end{cases}% \ln \mean{\mathcal{N}(E)}
    \label{eq:S_Boltz}
\end{align}
The canonical free energy follows from the Legendre transform
~\cite{Ruelle1969book,kardar2007}, $F(T)=\min_E\left[E-T\SMC(E)\right]$. For a saddle point $E_T$, the minimizing energy satisfies
\begin{align}
    \left.\frac{d\SMC(E)}{dE}\right|_{E_T}=\frac{1}{T}.
\end{align}
At sufficiently low temperatures, however, the minimum occurs at the lower entropy boundary, $E_T=-\EM$. Note that, for a single realization of the Hamiltonian, $\int_{E_1 - dE}^{E_1 + dE}\rho(x)dx = N^{-1}$ where $E_1$ is the ground state energy and $dE$ is much smaller than the mean level spacing. Thus, intuitively we expect that $ |\mean{E_1}| \approx \EM$, provided the DOS is self-averaging. $\EM$ defines the transition temperature
$\TM$ through
\begin{align}
    \left.\frac{d\SMC(E)}{dE}\right|_{-\EM}
    =NdE\left.\frac{d\mean{\rho(E)}}{dE}\right|_{-\EM}
    =\frac{1}{\TM}.
\end{align}
Consequently, the free energy takes the form
\begin{align}
    F(T)=
    \begin{cases}
        -\EM, & T\leq\TM,\\
        E_T-T\SMC(E_T), & T>\TM.
    \end{cases}
\end{align}
This piecewise behavior implies that $F(T)$ is non-analytic at $T = \TM$. Since an equilibrium phase transition is characterized by non-analytic behavior of the free energy, this non-analyticity signals an equilibrium phase transition at $T = \TM$. Hence, the shape of the average DOS, specifically the scaling at the edge of the spectrum determines a critical temperature, $\TM$ below which the ground state should dictate the thermodynamic properties of a system.

Apart from the microcanonical description in the canonical ensemble, the free energy is related to the partition function as $F=-T\ln Z(T)$, where the Boltzmann constant is set to unity. We then consider the heat capacity, which is defined in terms of the second derivative of the free energy as
\begin{align}
    \Capheat \equiv -T\dpd[2]{\FC}{T}
    = \frac{\meanT{E^2}-\meanT{E}^2}{T^2},
    \label{eq:Cv_def}
\end{align}
%---------------------------
%   Thermodynamic properties of beta ensemble (Heat capacity)
\begin{figure}[t]
\centering
\includegraphics[width=0.7\linewidth]{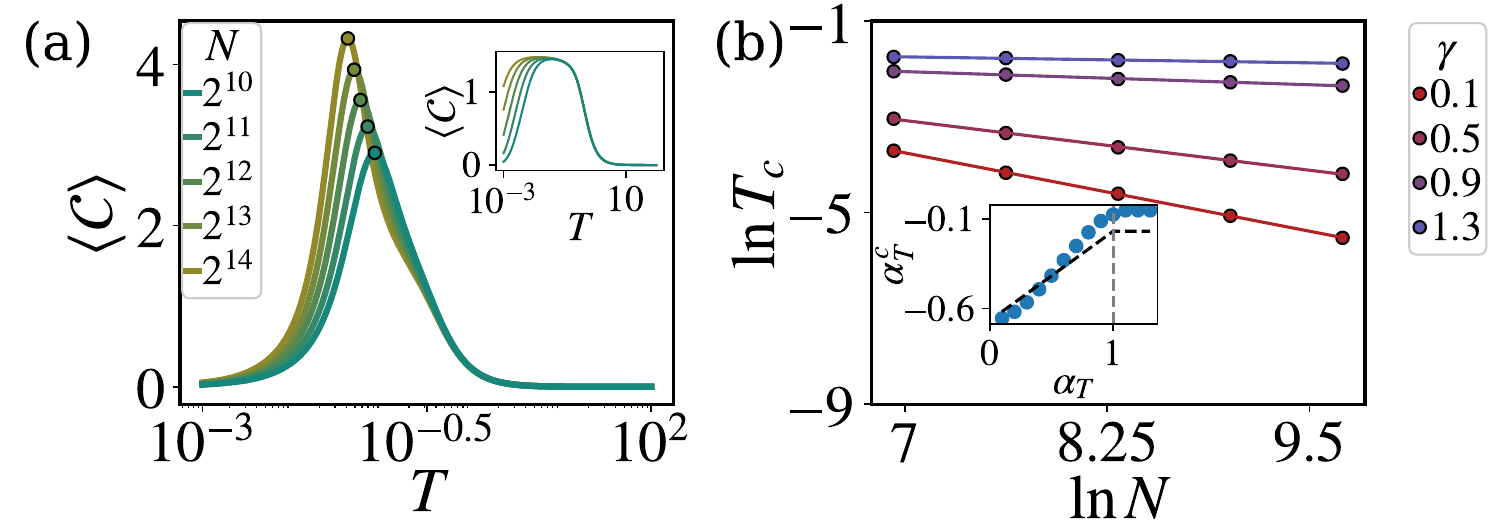}
\caption{disorder averaged heat capacity and the scaling of the critical temperature $\Tc$ in the \bte for $\sgE = 1/2$. (a)~disorder averaged heat capacity as a function of $T$ for different system sizes and $\gamma = 0.7$. Circular markers denote the critical temperature $\Tc$. Inset shows the heat capacity for $\gamma = 0$.
%===============
(b)~System size scaling of the critical temperature $\Tc$ for different values of $\gamma$. Inset shows the scaling exponent of $\Tc$ as a function of the system parameter $\gamma$. Black dashed line shows Eq.~\eqref{eq:alpha_T_c}.
}
\label{fig:C_Tc}
\end{figure}
%---------------------------
where $\meanT{E^k}\equiv\sum_n E_n^k p_n$ denotes the $k$th thermal moment of energy.
In Fig.~\ref{fig:C_Tc}(a), we show the temperature dependence of the ensemble-averaged heat capacity $\mean{\Capheat}$ for different system sizes and $\gamma = 0.7$, where we observe an increasing maximum height with increasing system size for $\gamma>0$. However for $\gamma \leq 0$ the heat capacity does not show any prominent maxima~\cite{Okuyama2020} as shown in the inset of Fig.~\ref{fig:C_Tc}(a).
In the thermodynamic limit, the free energy is independent of temperature for $T<\Tc$, so the heat capacity vanishes in this regime. It therefore has a jump at $T=\Tc$, consistent with a second-order phase transition. At finite system size, corrections to the free energy give a nonzero heat capacity for $T<\Tc$; these corrections decay very slowly with $N$~\cite{Derrida1981}. Thus, a heat-capacity maximum by itself does not establish a second-order phase transition, although its growth provides a useful finite-size indicator of $\Tc$ in the regime studied here. Nevertheless, the temperature $\Tc$ exhibits a power-law dependence on the system size, such that $\ln \Tc$ varies linearly with $\ln N$, which again justifies Eq.~\eqref{eq:alphaT_def}.
In Fig.~\ref{fig:C_Tc}(b) we show the system size scaling of $\Tc$ for different values of $\gamma$ $(>0)$ whereas in the inset we show the scaling exponent as a function of $\gamma$. In the canonical treatment, the entropy can be obtained from the canonical partition function as
\begin{align*}
    \mathcal{S}_{c}(T)= \mean{\ln Z(T)}+\frac{\meanT{E}}{T},
\end{align*}
%---------------------------
%   Thermodynamic properties of beta ensemble (S, Z, p1)
\begin{figure}[t]
\centering
\includegraphics[width=0.7\linewidth]{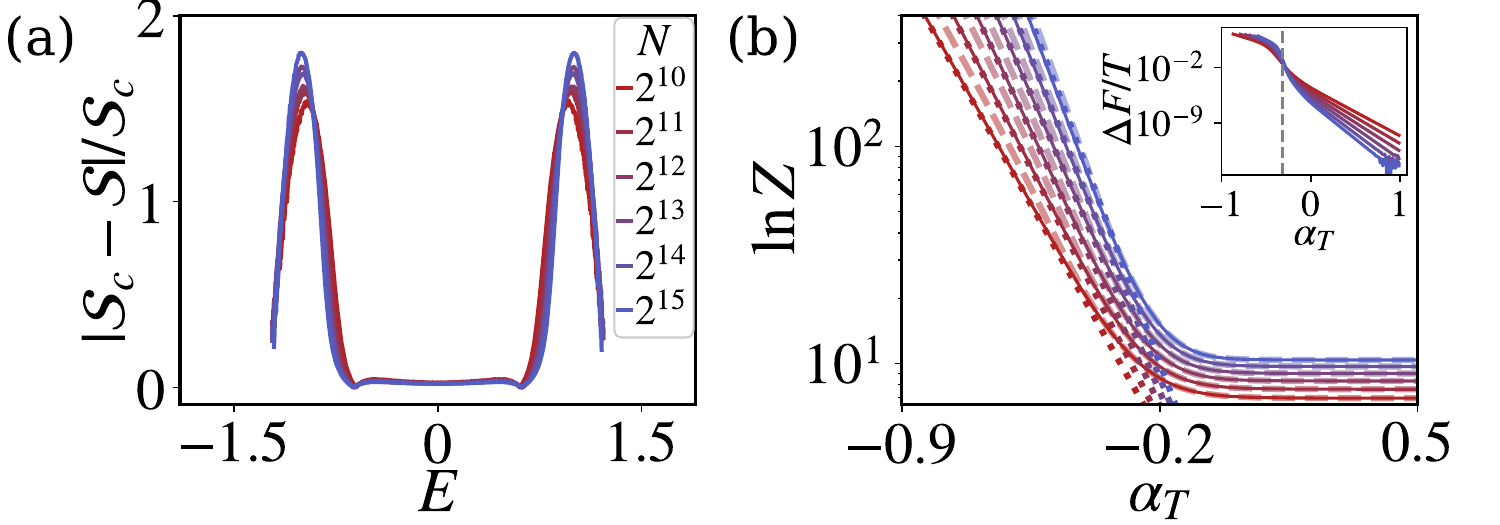}
\caption{Thermodynamic quantities of the \bte for $\sgE = 1/2$. (a)~Relative difference of the microcanonical and canonical entropy for different system sizes and $\gamma = 0.7$.
%===============
(b)~Logarithm of typical and average partition function shown via solid and dashed lines respectively for $\gamma = 0.7$ and different system sizes. The dotted lines show the empirical form given in Eq.~\eqref{eq:Z_typ_fit}. Inset shows $\Delta F / T$. Vertical line shows $\alpha_T^c$ according to Eq.~\eqref{eq:alpha_T_c}.
}
\label{fig:S_Z}
\end{figure}
%---------------------------
where \(\meanT{E}\) denotes the first thermal moment of the energy. We calculate the canonical entropy for each disorder realization and subsequently perform the disorder average. The resulting canonical entropy can then be compared with the microcanonical entropy introduced above in Eq.~\eqref{eq:S_Boltz} by evaluating \(\mathcal{S}(E)\) at the corresponding thermal energy \(E=\meanT{E}\). In Fig.~\ref{fig:S_Z}(a), we show the relative difference of the microcanonical and canonical entropies defined as $|\mathcal{S}_c - \mathcal{S}| / \mathcal{S}_c$ at a fixed value of $\gamma$ for different system sizes.
Note that, in the microcanonical picture
\begin{align}
    \mathcal{S}''(E)
    =
    \frac{d}{dE}\left(\frac{1}{T}\right)
    =
    -\frac{1}{T^2}\frac{dT}{dE},
\end{align}
and using $\Capheat=dE / dT$, we obtain $ \mathcal{S}''(E)=-\frac{1}{\Capheat T^2}$. Therefore, expanding the exponent around the saddle point $E_T$ gives,
\begin{align}
    \mathcal{S}(E)-\frac{E}{T}
    \simeq
    \mathcal{S}(E_T)-\frac{E_T}{T}
    -\frac{(E-E_T)^2}{2\Capheat T^2},
\end{align}
and hence the canonical partition function
\begin{align}
    Z(T) = \int dE e^{\mathcal{S}(E)-E/T}
    \simeq
    e^{\mathcal{S}(E_T)-E_T/T}
    \sqrt{2\pi \Capheat T^2}.
\end{align}
Using $F(T)=-T\ln Z(T)$, we obtain the canonical entropy,
\begin{align}
    \mathcal{S}_{\rm c}(T)
    =
    -\frac{\partial F}{\partial T}
    \simeq
    \mathcal{S}(E_T)
    +\frac{1}{2}\ln\left(2\pi \Capheat T^2\right),
\end{align}
where the leading contribution $\mathcal{S}(E_T)$ is the microcanonical entropy evaluated at the saddle-point energy, and the second term is the Gaussian fluctuation correction. Thus, while calculating the canonical entropy, we have subtracted the term $\frac{1}{2}\ln\left(2\pi \Capheat T^2\right)$, which appears as a correction to Eq.~\eqref{eq:S_Boltz}~\cite{kardar2007}.
We also look at the dependence of the typical and average partition function on the temperature parameter $\alpha_T$, as shown in Fig.~\ref{fig:S_Z}(b). An approximate form of the typical partition function at low temperatures can be given semi-analytically using the mean of the ground-state energy distribution.
We show Eq.~\eqref{eq:Z_typ_fit} along with the numerical data in Fig.~\ref{fig:S_Z}(b).
In the inset we show the quantity $\abs{\ln\mean{Z} - \mean{\ln Z}} \equiv \Delta F / T$. We observe that, at high $T > \Tc$ (low $T < \Tc$) temperatures, the difference decreases (increases) with system size. Thus, the system exhibits ensemble equivalence (inequivalence) at high (low) temperatures, corresponding to self-averaging (loss of self-averaging) in the respective regimes~\cite{Baldwin2020}.

%===========================================================
\section{Distribution of the ground state energy}\label{sec_appnd_E_dist}
%===========================================================
%---------------------------
% <E1> fit
\begin{figure}[t]
\centering
\includegraphics[width=0.4\linewidth]{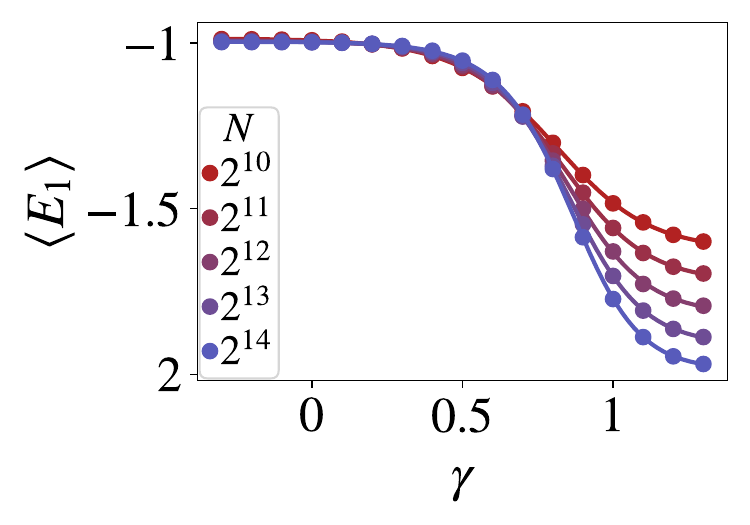}
\caption{$\mean{E_1}$ as a function of $\gamma$ for different system sizes. Circles show the numerical data while the solid lines show the fitted function given in Eq.~\eqref{eq:E1_mean_fit}.
}
\label{fig:E1_fit}
\end{figure}
%---------------------------
We propose the following fitting function for the mean ground-state energy, $\langle E_1\rangle$
\begin{align}
    \langle E_1\rangle
    &=-E_0(N)-A(N)
    \left[
    \frac{1}{2}
    \left (
    1+\tanh\left(
    \frac{\gamma-\gamma_c(N)}{\omega}
    \right)
    \right )
    \right]^p,
    \label{eq:E1_mean_fit}
\end{align}
with $\omega=0.25$ and $p=0.8$. The finite-size scaling gives
\begin{align}
    E_0(N)&\simeq 1-0.683N^{-0.578},\\
    A(N)&\simeq0.222N^{0.155},\\
    \gamma_c(N)&\simeq0.943-0.631N^{-0.267}.
\end{align}
The fitted form of Eq.~\eqref{eq:E1_mean_fit} along with the numerical data of $\mean{E_1}$ as a function of $\gamma$ for various $N$ are shown in Fig.~\ref{fig:E1_fit}.
Using the finite-size scaling of the fit parameters, the typical partition function at low temperatures is then approximated as
\begin{align}
    \ln Z_{\mathrm{typ}}(T) \approx - \frac{\mean{E_1}}{T}
    &\simeq
    \frac{E_0(N)+A(N)
    \left[
    \frac{1}{2}
    \left (
    1+\tanh\left(
    \frac{\gamma-\gamma_c(N)}{0.25}
    \right)
    \right )
    \right]^{0.8}}{T}.
    \label{eq:Z_typ_fit}
\end{align}
%===========================================================
\section{Analytical expression of the finite temperature SFF}\label{sec_appnd_SFF_ana}
%===========================================================
To get an approximate form the finite temperature SFF, it is convenient to recast Eq.~\eqref{eq:SFF_T_def} into the equivalent form
\begin{align}
	\sff{t;T}
	=
	\int dE\,
	e^{-iEt}
	\sum_{m,n}
	p_m p_n\,
	\delta(E-E_m+E_n),
	\label{eq:SFF_T_def_3}
\end{align}
where $p_n$ is the Gibbs occupation probability associated with the energy level $E_n$. This Equation explicitly contains both equal-level and distinct-level contributions. Averaging over disorder realizations gives
\begin{align}
	\mean{\sff{t;T}}
	=
	\sffbar
	+
	\mean{
	\int dE\,
	e^{-iEt}
	\sum_{m\neq n}
	p_m p_n\,
	\delta(E-E_m+E_n)
	}.
	\label{eq:SFF_T_step1}
\end{align}
{\bfseries Saturation value of the SFF:}
The first term in Eq.~\eqref{eq:SFF_T_step1} is the stationary contribution,
\begin{align}
	\sffbar
	&=
	\mean{
	\int dE\,
	e^{-iEt}
	\sum_m
	p_m^2\,
	\delta(E)
	}
	=
	\mean{\sum_m p_m^2}
	=
	\mean{\frac{\nrm(T/2)}{\nrm(T)^2}},
	\label{eq:SFF_T_bar}
\end{align}
representing the long-time saturation value of the SFF. In the high-temperature regime, the long-time saturation value of the SFF can be simplified within the annealed approximation as
\begin{align}
	\sffbar
	\simeq
	\frac{\mean{\nrm(T/2)}}{\mean{\nrm(T)}^2}.
\end{align}
We next consider the second term of Eq.~\eqref{eq:SFF_T_step1} that incorporates correlations between different energy levels. By explicitly substituting the Gibbs probabilities, it can be written as
\begin{align}
	\begin{split}
		&
		\mean{
		\int dE\,e^{-iEt}
		\sum_{m\neq n}
		p_m p_n
		\delta(E-E_m+E_n)
		}
		\\
		=&
		\int dE\,e^{-iEt}
		\int d\vec{E}\,
		\prob{\vec{E}}
		\sum_{m\neq n}
		\frac{e^{-(E_m+E_n)/T}}
		{\nrm(T)^2}
		\delta(E-E_m+E_n)
		\\
		=&
		\int d\vec{E}\,
		\prob{\vec{E}}
		\sum_{m\neq n}
		\frac{\exp\!\left[
		-\frac{E_m+E_n}{T}
		-i(E_m-E_n)t
		\right]}
		{\nrm(T)^2}
		\\
		=&
		\int d\vec{E}\,
		\prob{\vec{E}}
		\sum_{m\neq n}
		\frac{\exp\!\left[
		-\frac{E_{mn}}{T}
		-i\Omega_{mn}t
		\right]}
		{\nrm(T)^2},
		\qquad
		\begin{matrix}
		E_{mn}\equiv E_m+E_n,\\
		\Omega_{mn}\equiv E_m-E_n.
		\end{matrix}
	\end{split}
	\label{eq:SFF_T_term2a}
\end{align}
where $\prob{\vec{E}}$ is the joint probability density function (JPDF) of the complete set of eigenvalues, $\vec{E} = \left(E_1,\ldots,E_N\right)$.
An exact evaluation of this expression is generally not feasible because the partition function in the denominator depends on the entire spectrum. This difficulty can be overcome at $\alpha_T> \alpha_T^c$, by employing the annealed approximation, 
\begin{align}
	\mean{\sff{t;T}}
	\simeq
	\frac{\mean{|\nrm(T,it)|^2}}
	{\mean{\nrm(T)}^2}.
\end{align}

In order to evaluate the integral we need the two-level correlation function $\rho^{(2)}(E_m,E_n)$ that describes the joint probability density of finding two distinct eigenvalues at energies $E_m$ and $E_n$. Upon ensemble averaging, in terms of the JPDF of the eigenvalues, this is given by
\begin{align}
    \mean{\rho^{(2)}(E_m,E_n)}
    &=\int \prod_{i=1}^{N}dE_i\,
    \prob{\vec{E}}
    \sum_{i\neq j}
    \delta(E_m-E_i)\delta(E_n-E_j).
\end{align}
Next, the two-level correlation function can be decomposed into a disconnected contribution, given by the product of the one-level densities defined in Eq.~\eqref{eq:rho_E}, and a connected contribution described by the two-level cluster function $T_2(E_m,E_n)$, which quantifies the deviation from the independent-eigenvalue contribution and thus captures spectral correlations:
\begin{align}
    \mean{\rho^{(2)}(E_m,E_n)}
    =
    N^2
    \mean{\rho(E_m)}
    \mean{\rho(E_n)}
    -
    T_2(E_m,E_n),
\end{align}
then ignoring the ordering constraint among eigenvalues leads to
\begin{align}
	\begin{split}
		&
		\frac{1}{\mean{\nrm(T)}^2}
		\int d\vec{E}\,
		\prob{\vec{E}}
		\sum_{m\neq n}
		\exp\!\left(
		-\frac{E_{mn}}{T}
		-i\Omega_{mn}t
		\right) = \mean{\sffdc{t; T}} + \mean{\sffc{t; T}},
	\end{split}
	\label{eq:SFF_T_term2}
\end{align}
where the first term is the disconnected part of the SFF and the second term is the connected part of the SFF.  \\
{\bfseries Disconnected part of the SFF:}
The disconnected part is
\begin{align}
    \mean{\sffdc{t; T}} & \equiv \frac{1}{\mean{\nrm(T)}^2}
		\frac{(N-2)!}{N!}
		\sum_{m\neq n}
		\int dE_m\,dE_n\,
		N^2
		\mean{\rho(E_m)}
		\mean{\rho(E_n)}
		e^{-E_{mn}/T-i\Omega_{mn}t} \\
    &\approx\;
		\frac{N^2}{\mean{\nrm(T)}^2}
		\int dE_m\,dE_n\,
		\mean{\rho(E_m)}
		\mean{\rho(E_n)}
		e^{-E_{mn}/T-i\Omega_{mn}t}.
\end{align}
This term contains no information about level correlations and depends exclusively on the thermally weighted DOS. It can be recast as
\begin{align}
	\begin{split}
		&
		\int dE_m\,dE_n\,
		e^{-i(E_m-E_n)t}
		p_m p_n
		N^2
		\mean{\rho(E_m)}
		\mean{\rho(E_n)}
		\\
		=&
		\int dE_m\,dE_n\,
		e^{-i(E_m-E_n)t}
		\mean{\rho_T(E_m)}
		\mean{\rho_T(E_n)}
		\\
		=&
		\left|
		\int dE\,
		\mean{\rho_T(E)}
		e^{-iEt}
		\right|^2,
	\end{split}
	\label{eq:SFF_T_slope}
\end{align}
where $\rho_T(E)$ is the thermally weighted DOS defined in Eq.~\eqref{eq:rho_T_CGS_def}.
The disconnected part therefore reflects the smooth decay of the SFF arising from the overall spectral envelope.\\
{\bfseries Connected part of the SFF:}
The cluster function is approximated as
\begin{align}
    T_2(E_m,E_n)
    \simeq N ^2
    \mean{\rho\!\left(
    \frac{E_m+E_n}{2}
    \right)}^2
    Y_2\!\left[ N
    \mean{\rho\!\left(
    \frac{E_m+E_n}{2}
    \right)}
    (E_m-E_n)
    \right],
    \label{eq:T2_local}
\end{align}
where $Y_2$ is the universal unfolded two-level cluster function. The connected contribution to the spectral form factor is
\begin{align}
    \mean{\sffc{t;T}}
    =
    -
    \frac{1}{\mean{\nrm(T)}^2}
    \frac{(N-2)!}{N!}
    \sum_{m\neq n}
    \int dE_m\,dE_n\,
    T_2(E_m,E_n)
    e^{-E_{mn}/T}
    e^{-i\Omega_{mn}t},
    \end{align}
    where
    \begin{align}
    E_{mn}=E_m+E_n,
    \qquad
    \Omega_{mn}=E_m-E_n.
\end{align}
Using
\[
\frac{(N-2)!}{N!}\sum_{m\neq n}=1,
\]
we obtain
\begin{align}
    \mean{\sffc{t;T}}
    \simeq
    -
    \frac{1}{\mean{\nrm(T)}^2}
    \int dE_m\,dE_n\,
    T_2(E_m,E_n)
    e^{-E_{mn}/T}
    e^{-i\Omega_{mn}t}.
\end{align}
Introducing the centre and relative coordinates
\begin{align}
    E=\frac{E_m+E_n}{2},
    \qquad
    \Omega=E_m-E_n,
\end{align}
with unit Jacobian
\begin{align}
    dE_m\,dE_n=dE\,d\Omega,
\end{align}
gives
\begin{align}
    \begin{split}
    \mean{\sffc{t;T}}
    =
    -
    \frac{1}{\mean{\nrm(T)}^2}
    \int dE\,
    N^2\mean{\rho(E)}^2
    e^{-2E/T}
    \int d\Omega\,
    Y_2\!\bigl(\mean{\rho(E)}\Omega\bigr)
    e^{-i\Omega t}.
    \end{split}
    \label{eq:SFF_local}
\end{align}
We now locally unfold the spectrum by defining
\begin{align}
\omega=N\mean{\rho(E)}\Omega,
\qquad
d\Omega=\frac{d\omega}{N\mean{\rho(E)}}.
\end{align}
Substituting this into Eq.~(\ref{eq:SFF_local}) yields
\begin{align}
    \begin{split}
    \mean{\sffc{t;T}}
    =
    -
    \frac{1}{\mean{\nrm(T)}^2}
    \int N dE\,
    \mean{\rho(E)}
    e^{-2E/T}
    \int d\omega\,
    Y_2(\omega)
    \exp\!\left(
    -i\frac{\omega t}{N\mean{\rho(E)}}
    \right).
    \end{split}
\end{align}
Within the local bulk approximation, the DOS varies only slowly over the thermal window and may therefore be replaced by its
average bulk value,
\[
\mean{\rho(E)}\simeq\bar\rho.
\]
The $\omega$-integral then becomes independent of the centre coordinate,
\begin{align}
    \int d\omega\,
    Y_2(\omega)
    e^{-i\omega \tau}
    =
    b_2(\tau),
\end{align}
where
\begin{align}
    \tau=\frac{t}{t_H},
    \qquad
    t_H=\frac{1}{\Delta},
    \qquad
    \Delta=\frac{1}{N\bar\rho},
\end{align}
is the Heisenberg time. The connected contribution therefore factorizes as
\begin{align}
    \begin{split}
    \mean{\sffc{t;T}}
    &=
    -
    \frac{b_2(\tau)}
    {\mean{\nrm(T)}^2}
    \int N dE\,
    \mean{\rho(E)}
    e^{-2E/T}
    \\
    &=
    -
    \frac{\mean{\nrm(T/2)}}
    {\mean{\nrm(T)}^2}
    b_2(\tau)
    \\
    &=
    -
    \sffbar\,b_2(\tau),
    \end{split}
    \label{eq:SFF_T_ramp}
\end{align}
where
\begin{align}
    \sffbar
    =
    \frac{\mean{\nrm(T/2)}}
    {\mean{\nrm(T)}^2}.
\end{align}\\

Finally, combining Eqs.~\eqref{eq:SFF_T_slope}, \eqref{eq:SFF_T_ramp}, and \eqref{eq:SFF_T_bar} yields
\begin{align}
	\mean{\sff{t;T}}
	\approx
	\left|
	\int dE\,
	\mean{\rho_T(E)}
	e^{-iEt}
	\right|^2
	+
	\sffbar
	\left[
	1-b_2(\tau)
	\right],
	\label{eq:SFF_T_final}
\end{align}
which naturally separates the two mechanisms governing the finite-temperature SFF. The first term is entirely determined by the thermally weighted DOS and describes the nonuniversal decay of the SFF, whereas the second term contains the universal contribution arising from spectral correlations through the two-level form factor. Note that, we can replace the DOS by its
average bulk value only in the regime $\alpha_T\geq \alpha_T^q$. Thus the approximation of Eq.~\eqref{eq:SFF_T_final}, involving the two-level form factor is only valid in the annealed phase.

%===========================================================
\section{Validity of the excited-state sum}\label{sec_appnd_ex_sum}
%===========================================================
%---------------------------
% D_kbar for large gamma
\begin{figure}[t]
\centering
\includegraphics[width=0.4\linewidth]{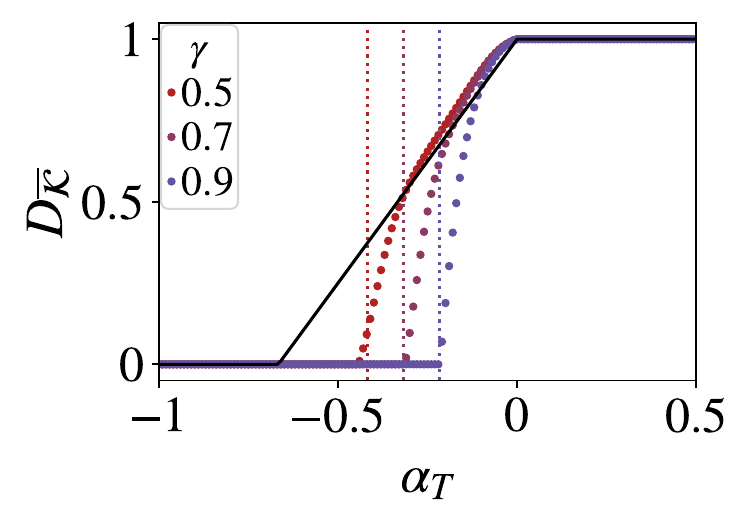}
\caption{$D_{\sffbar}$ vs $\alpha_T$ for different values of $\gamma$, away from $\gamma = 0$. Vertical dotted lines marks $\alpha_T^c$ from Eq.~\eqref{eq:alpha_T_c} and the black solid line shows the analytical expression of Eq.~\eqref{eq:Dp1_Dkbar_ana}.}
\label{fig:Dkbar_app}
\end{figure}
%---------------------------
We examine the conditions under which the excited-state sum in Eq.~\eqref{eq:excited_state_sum} remains bounded in the thermodynamic limit. A breakdown of the bounded-sum assumption would require the excited-state Boltzmann weights to decay sufficiently slowly that their sum diverges over a growing range of levels. For example, weights decaying as slowly as \(n^{-1}\) up to \(n\sim N^a\), with \(a>0\), would produce a logarithmic divergence. More generally, weights decaying as \(n^{-1+\epsilon}\), with \(\epsilon>0\), would produce a power-law divergence.

At \(T_0=T_qN^{-c}\), such a slow decay would require excitation energies growing approximately as
\begin{equation}
    E_n-E_1\lesssim T_0\ln n
    \sim \sigma_E N^{-c}\ln n,
\end{equation}
where \(T_q\) is of order \(\sigma_E\). Equivalently, the DOS would have to increase exponentially from the spectral edge.

Such behaviour is not expected for the semicircular DOS, which vanishes as a square root at the edge. To see this explicitly, consider the excited-state sum
\begin{equation}
    S(T)=\sum_{n>1}e^{-(E_n-E_1)/T}.
\end{equation}
Using \(E_1\simeq-2\sigma_E\) and the semicircular DOS, in the continuum edge regime \(E_2-E_1\ll T\ll 2\sigma_E\), we obtain
\begin{align}
    S(T)
    &\simeq N\int_0^\infty dx\,
    \frac{\sqrt{x}}{\pi\sigma_E^{3/2}}e^{-x/T}\\
    &=\frac{N}{2\sqrt{\pi}}
    \left(\frac{T}{\sigma_E}\right)^{3/2}.
\end{align}
Similarly,
\begin{equation}
    S(T/2)\simeq 2^{-3/2}S(T).
\end{equation}
Consequently, Eq.~\eqref{eq:Kbar_p1_ratio} becomes
\begin{equation}
    \frac{\overline{\mathcal K}}{p_1}
    \simeq
    \frac{1+2^{-3/2}S(T)}{1+S(T)}.
\end{equation}
This ratio remains finite and nonzero, approaching \(2^{-3/2}\) when \(S(T)\gg1\), and unity when \(S(T)\ll1\). Thus, even when the excited-state sum itself grows with system size, its growth does not lead to a breakdown of the equality of the scaling exponents of \(\overline{\mathcal K}\) and \(p_1\) for the semicircular DOS. Fig.~\ref{fig:Dkbar_app} shows the deviation of $D_{\sffbar} $ from Eq.~\eqref{eq:Dp1_Dkbar_ana} for $\gamma$ values away from $0$.
%===========================================================
\section{Details of the FTLM calculation}\label{sec_appnd_FTLM}
%===========================================================
%---------------------------
% FTLM
\begin{figure}[t]
\centering
\includegraphics[width=0.7\linewidth]{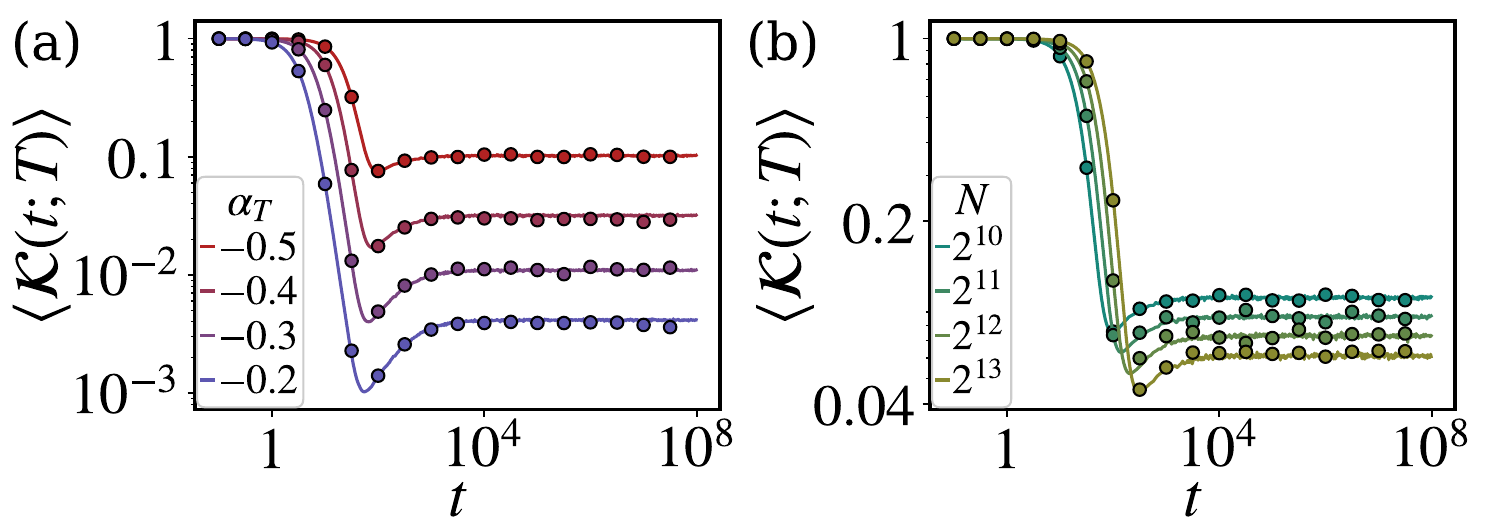}
\caption{(a)~Finite temperature SFF for $\gamma = 0.2$, $N = 1024$ and different values of $\alpha_T$. The circular markers are obtained using FTLM whereas the solid line shows the SFF obtained using exact diagonalization method.
%===============
(b)~Finite temperature SFF for $\gamma = 0.2$, $\alpha_T = -0.5$ and different system sizes. 
}
\label{fig:FTLM_app}
\end{figure}
%---------------------------
In the FTLM calculations, for each disorder realization, we use $N_v=128$ initial random vectors in Eq.~\eqref{eq:ZT_FTLM} and then average the resulting quantity over different realizations.

To estimate the minimum number of Lanczos iterations $m_\mathrm{opt}$ required to obtain reasonable agreement with exact diagonalization, we calculate the relative error $\delta \sffbar_\mathrm{FTLM} = \frac{|\sffbar - \sffbar_\mathrm{FTLM}|}{\sffbar}$ between the long-time average of the FTLM SFF and the corresponding saturation value $\sffbar$ obtained from exact diagonalization. We choose a relative-error threshold of $10\%$ and define $m_\mathrm{opt}$ as the minimum number of Lanczos iterations for which $\delta \sffbar$ falls below it. As $\alpha_T$ increases, $m_\mathrm{opt}$ shifts towards $\mathcal{O}(N)$ because the Gibbs distribution becomes increasingly uniform.

Figure~\ref{fig:FTLM_app}(a) compares the SFF obtained using FTLM with $m_\mathrm{opt}$ Lanczos iterations against the exact diagonalization results for different values of $\alpha_T$, while Fig.~\ref{fig:FTLM_app}(b) shows the corresponding comparison for different system sizes at fixed $\alpha_T$ and $\gamma$. This figure emphasizes the use of $m \ll N$ to estimate the SFF at low temperatures.

%===========================================================
\section{Mean-energy derivation}\label{sec_appnd_meanE}
%===========================================================
Using the annealed approximation, the ensemble-averaged thermally weighted DOS is given in Eq.~\eqref{eq:rho_T_approx} and using that the ensemble-averaged thermal energy, Eq.~\eqref{eq:mean_E}, can then be expressed as
\begin{align}
    \overline{\langle E\rangle_T}
    \simeq \int_{-\infty}^{\infty}
    E\langle\rho_T(E)\rangle\,dE
    =\frac{1}{2\pi\sigma_E T I_1(2\sigma_E/T)}
    \int_{-2\sigma_E}^{2\sigma_E}
    E\sqrt{4\sigma_E^2-E^2}\,e^{-E/T}\,dE.
\end{align}
We can write the integral as
\begin{align}
    \int_{-2\sigma_E}^{2\sigma_E}
    E\sqrt{4\sigma_E^2-E^2}\,e^{-E/T}\,dE
    =-\frac{\partial}{\partial(1/T)}
    \left[2\pi\sigma_E T I_1\left(\frac{2\sigma_E}{T}\right)\right],
\end{align}
and then using the identity $\frac{dI_1(x)}{dx}=I_0(x)-\frac{I_1(x)}{x}$, we obtain
\begin{align}
    \overline{\langle E\rangle_T}
    \simeq 2T-2\sigma_E
    \frac{I_0(2\sigma_E/T)}{I_1(2\sigma_E/T)}.
\end{align}
Finally, expanding the modified Bessel functions at $T\gg\sigma_E$ gives
\begin{align}
    \overline{\langle E\rangle_T}
    \simeq -\frac{\sigma_E^2}{T}
    +\frac{\sigma_E^4}{6T^3}
    +\mathcal{O}\left(\frac{\sigma_E^6}{T^5}\right).
\end{align}
%=======================================
\bibliography{references}

\begin{thebibliography}{72}
\bibitem [{Not()}]{Note}%
  \BibitemOpen
  \href@noop {} {}\bibinfo {note} {Please note that in
  Ref.~\onlinecite{Roy2025} a different convention for the normalization of the
  spectrum was used, where the global mean level spacing was $\delta \sim
  \mathcal{O}(1)$. Accordingly, the critical energy scale was $E_c \sim
  N^\gamma$, and the timescale exponents $\mu_{\mathrm R}$,
  $\mu_{\mathrm{dip}}$, and $\mu_{\mathrm{Zeno}}$ were each shifted by $-1$
  relative to the values used here.}\BibitemShut {Stop}%
\end{thebibliography}
\end{document}